\documentstyle[epsf,aps]{revtex} 		
\newcommand{\bfi}[1]{\mbox{\boldmath $#1$}}
\newcommand{\Lower}[1]{\smash{\lower 1.5ex \hbox{#1}}}
\newcommand{\LN}{$\Lambda N$ }
\newcommand{\LL}{$\Lambda\Lambda$ }
\newcommand{\HeVL}{$_\Lambda^5$He}
\newcommand{\HVLL}{$_{\Lambda\Lambda}^{\ \ 5}$H}

\newcommand{\HeVILL}{$_{\Lambda\Lambda}^{\ \ 6}$He}

\title{Study of Light $\Lambda$- and $\Lambda\Lambda$-Hypernuclei 
with the Stochastic Variational Method and 
Effective $\Lambda N$ Potentials}

\author{Hidekatsu {\sc Nemura}$^1$, Yasuyuki {\sc Suzuki}$^2$, 
Yoshikazu {\sc Fujiwara}$^3$ and Choki {\sc Nakamoto}$^{4}$} 
\address{ 					
$^1$Graduate School of Science and Technology, Niigata University, 
Niigata 950-2181, Japan\\
$^2$Department of Physics, Niigata University, Niigata 950-2181, Japan\\
$^3$Department of Physics, Kyoto University, Kyoto 606-8502, Japan \\
$^4$Suzuka National College of Technology, Suzuka, Mie 510-0294, Japan}

\date{\today}

\begin{document} 				

\maketitle 					

\begin{abstract} 				
We first determine the \LN $S$-wave phase shifts so as to  
reproduce the experimental $\Lambda$ separation energies 
 of $A=3, \,4$ $\Lambda$-hypernuclei ($_\Lambda^3$H, $_\Lambda^4$H, 
 $_\Lambda^4$He, $_\Lambda^4$H$^\ast$, and $_\Lambda^4$He$^\ast$), 
and then construct three phase-equivalent \LN potentials with 
different central repulsion. By the stochastic variational method 
with correlated Gaussian basis we perform an extensive calculation 
of {\it ab initio} type for the hypernuclei of up to $A$=6. 
The binding energies and the sizes of the $\Lambda$-hypernuclei 
are very insensitive to the type of the phase-equivalent \LN 
potentials. We use two different \LL potentials which both reproduce 
$\Delta B_{\Lambda\Lambda}(_{\Lambda\Lambda}^{\ \ 6}$He) reasonably well. 
Any combination of these \LN and \LL potentials predicts hitherto 
undiscovered particle-stable bound states, $_{\Lambda\Lambda}^{\ \ 4}$H, 
$_{\Lambda\Lambda}^{\ \ 5}$H and $_{\Lambda\Lambda}^{\ \ 5}$He: 
Predicted values of $B_{\Lambda\Lambda}$ are about 0.4, 5.5 and 6.3 MeV, 
respectively. The binding energy of $_{\Lambda\Lambda}^{\ \ 4}$H is 
so small that its possibility crucially depends on the strength 
of the \LL interaction. The binding energies of both $^5_{\Lambda}$He and 
$_{\Lambda\Lambda}^{\ \ 6}$He are calculated to be strongly 
overbound compared to experiment. In relation to this well-known anomaly 
we examine the effect of the quark substructure of $N$ and 
$\Lambda$ on their binding energies. The effect is negligible if the 
baryon size in which three quarks are confined 
is smaller than 0.6 fm, but becomes appreciable, particularly 
in $_{\Lambda\Lambda}^{\ \ 6}$He, if the size is taken to be as large 
as 0.7 fm. We discuss the extent to which the nucleon subsystem 
in the hypernuclei changes by the addition of $\Lambda$ particles. 
The charge symmetry breaking of the $\Lambda N$ potential is 
phenomenologically determined and concluded to be weakly spin-dependent.
\end{abstract} 					


\section{Introduction}

Strangeness or hypernuclear physics has been attracting increasing 
interest in recent years\cite{proc}. 
Experimental progress on both a production of hypernuclei 
with strangeness $S=-1$ and $-2$ and an analysis of their 
decay modes will bring rich information on the baryon dynamics 
involving strange particles. It may be particularly interesting to 
make clear to what extent the properties of a strange particle 
in nuclear medium change from those in a free space.  

The knowledge of the hyperon-nucleon ($YN$) interaction is crucially 
important for a further development of the hypernuclear physics. 
Because of the limited data on $\Lambda$-nucleon ($\Lambda N$) 
scatterings, no reliable phase shift analysis has been 
performed yet. 
Even the ${\bfi \sigma}_\Lambda\!\cdot\!{\bfi \sigma}_N$ part 
of the \LN interaction has not been determined precisely. 
Though one-boson-exchange (OBE) models\cite{ND77,NF79} or quark cluster 
models\cite{FSS96} 
for $YN$ interactions have 
produced useful results on the interactions and have shown some 
predictive power, 
the behavior of the $^1S_0$ or the $^3S_1$ phase shift 
is not always the same in these models.   
At present it is not possible to pin down 
realistic potentials which can be used in {\it ab initio} 
hypernuclear structure calculations. 

In view of this circumstance, hypernuclear spectroscopic study will be 
aided by modeling effective $\Lambda N$ and $\Lambda \Lambda$ potentials. 
One may determine the potentials phenomenologically so as 
to reproduce the binding energies 
of light $\Lambda$- and $\Lambda\Lambda$-hypernuclei before theoretically 
sound, realistic potentials become available. 
In fact, the early work\cite{DHT72} performed more than a quarter of 
a century ago with the effective $\Lambda N$ potential  
demonstrated that the observed binding energy of $_\Lambda^5$He is 
anomalously small. 
This problem still remains an enigma\cite{GIB95} despite of several attempts 
at resolving this anomaly by the mechanism of e.g., 
tensor forces\cite{Shi84} or $\Lambda N$-$\Sigma N$ couplings. 
Since the excited states 
of the $A\!=\!4$ isodoublets ($_\Lambda^4$H$^\ast$ 
and $_\Lambda^4$He$^\ast$) had 
not been well established in those days, 
it will be worthwhile to redetermine effective \LN potentials 
which include the charge symmetry breaking (CSB) effect. 

The information on effective \LN interactions 
becomes available together with the progress in technique for 
a precise solution of few-body systems. A detailed, 
systematic analysis of the binding mechanism of light hypernuclei 
may make it possible to determine the $S$-wave strength of the 
\LN interaction. As is well-known, the binding energies of the 
$A\!=\!3,\,4$ hypernuclei strongly depend on the relative 
strength of attraction between $^1S_0$ and $^3S_1$ states. 
If the $^1S_0$ attraction is only slightly stronger than 
the $^3S_1$ one as in the case of the Nijmegen model F\cite{NF79}(NF) 
or Dalitz {\it et al.}\cite{DHT72} potential, 
we would have not only the ground state 
$(J^\pi={1\over 2}^+)$ but also a particle-stable excited 
state ($J^\pi={3\over 2}^+$) of $_\Lambda^3$H. The latter 
has not experimentally been confirmed yet. 
This type of potentials predicts a too small energy difference of the two 
states (0$^+$ and 1$^+$) for the $A=4$ isodoublets.
On the contrary, one of the versions of the quark cluster model, 
FSS\cite{FSS96},  
gives a very strong 
attraction in $^1S_0$ and leads to the overbinding of $_\Lambda^3$H 
as well as a too large energy difference for the two states in 
the $A=4$ system. 
Recently the simplest three baryon system $YNN$ with $S=-1$  
has been studied\cite{Miy95} in the Faddeev method, which shows that 
only the Nijmegen soft core (NSC89) potential\cite{NSC89}, among others,  
reproduces the experimental $B_\Lambda(_\Lambda^3\mbox{H})$ value. 
The work has also shown clearly the importance for the 
$\Lambda N$-$\Sigma N$ conversion and its $^3S_1$-$^3D_1$ tensor 
coupling for producing the bound state of $_\Lambda^3$H. 
The new version of the Nijmegen soft core model (NSC97\cite{NSC97}) has been  
constructed on the basis of these theoretical developments. 

The purpose of the present study is threefold: 
First we determine phenomenologically 
such effective $\Lambda N$ central potentials 
that reproduce the binding energies of $s$-shell hypernuclei 
($_\Lambda^3$H, $_\Lambda^4$H, $_\Lambda^4$He, $_\Lambda^4$H$^*$, 
$_\Lambda^4$He$^*$). Secondly, we explore the possibility of 
whether $\Lambda\Lambda$-hypernuclei with $A\!=\!4,\,5$ form bound states 
or not by employing the available $\Lambda\Lambda$ potentials. 
This will be useful for experimental search for these 
$\Lambda\Lambda$-hypernuclei. A confirmation of the 
lightest $\Lambda\Lambda$-hypernuclei 
as well as the H-particle will give us useful information on the 
strength of the $\Lambda\Lambda$ interaction. 
Thirdly, we investigate to what extent the quark substructure 
of the baryons plays a role in reducing the $\Lambda$ separation 
energy of the hypernuclei with $A\ge 5$. 

The third point mentioned above is expected to reveal, through hypernuclear 
studies, a unique role of the quark substructure of baryons. 
The binding energies of the $\Lambda$-hypernuclei have so far 
been calculated on the 
premise that they ought to be understood provided two- and three-body 
baryonic interactions are precisely known. 
This seems quite natural because the $\Lambda$-particle can be 
distinguished from the nucleon at the baryon level. 
However, if we take into account the quark substructure of the baryon, 
the Pauli principle acting at the underlying quark level produces certain 
constraints on the dynamical motion of the hypernuclear system, because 
both $N$ and $\Lambda$ contain $u$ and $d$ quarks. 
This quark Pauli effect has recently been studied in Refs. 
\cite{Nem99} and \cite{Suz99} as a function of the baryon size $b$ 
in which three quarks are confined. 
The reduction of the binding energy 
of $_\Lambda^5$He and $_{\Lambda\Lambda}^{\ \ 6}$He is found to occur 
at $b \ge 0.6$ fm. 

It is of vital importance for achieving the above objectives that we can 
obtain a precise solution of few-body problems. 
A fair description of the correlated motion is an 
essential ingredient to get accurate solutions. 
As tested in a variety of examples in atomic and nuclear few-body 
problems \cite{VS95,SVM98}, the correlated Gaussian (CG) basis leads to 
a virtually exact solution. 
This basis explicitly describes the correlated 
motion of the constituent particles, and its simple form makes it 
possible to obtain matrix elements analytically. 
We therefore use a variational trial function given as a combination of 
the CGs, and increase the basis dimension one by one till a 
practical convergence is attained. The optimization of the 
nonlinear parameters contained in the basis functions is performed by 
selecting the best among a number of candidates randomly chosen.  
This gambling procedure is called the stochastic variational method (SVM) 
\cite{VS95,SVM98,Kuk}. 
   
We will focus on central components, both 
spin-independent and spin-dependent, of the potential, so the 
systems considered here are 
limited to those hypernuclei whose total orbital 
angular momentum is well approximated by $L=0$. 
We presume that the effects of tensor forces, three-body forces and 
$\Lambda$-$\Sigma$ conversion etc. are effectively replaced 
by a central force in fitting the binding energies of the $s$-shell 
hypernuclei. The hypernuclei with $A\le 6$ are treated, without 
any compromise, as an $A$-baryon system 
and their energies are calculated by fully taking into account the 
dynamics of the constituent baryons. 

We present in sect.\,2 the formalism of the present study. 
The effective $\Lambda N$ potential is determined in 
subsect.\,2.1 by starting 
from potentials phase equivalent to the Nijmegen model\cite{NF79} or 
the quark cluster model\cite{FSS96}. 
The $\Lambda\Lambda$ potentials used in the present study are 
briefly mentioned in this subsection. 
The variational trial function with the CGs 
is given in subsect.\,2.2. The optimization of the parameters 
of the trial function is carried out as described in subsect.\,2.3. 
Results of calculation are given in sect.\,3. 
The $\Lambda$-hypernuclei with $A\le 5$ and 
$\Lambda\Lambda$-hypernuclei with $A\le 6$ are discussed 
in subsects.\,3.1 and 3.2, respectively. 
The $\Lambda$ density distribution and the $\Lambda \Lambda$-correlation 
function are calculated in subsect.\,3.3 in order to exhibit the 
structure of the hypernuclei. We also analyze the extent to which 
the nuclear core changes by adding $\Lambda$ particles. 
The CSB term of the $\Lambda N$ potential is discussed in 
subsect.\,3.4 and its strength is determined phenomenologically. 
We consider in subsect.\,4.1 a special configuration 
constrained by the Pauli principle at the quark level. 
Results for $_\Lambda^5$He, $_\Lambda^5$H and 
$_{\Lambda\Lambda}^{\ \ 6}$He 
are presented in subsect.\,4.2. Summary is given in sect.\,5. 
Some formulas for the matrix elements of the CGs 
are collected in appendix. 

\section{The Formalism}
\subsection{Potentials}
As mentioned in the introduction, our aim in this subsection 
is to determine effective \LN 
central forces that presumably include various effects such as 
the tensor force, the three-body force and so on. 
The Minnesota potential\cite{Min77} is such a type of central 
potential widely employed as the $NN$ potential. The Minnesota 
potential is given in a form 
\begin{eqnarray}
V= \left\{ V_R + \frac{1}{2} (1+P^{\sigma}) V_t 
	+ \frac{1}{2} (1-P^{\sigma}) V_s \right\}
	\left\{ \frac{1}{2} u + \frac{1}{2} (2 - u) P^{r} \right\}, 
\label{MINNPOT} \\ 
V_{R} = V_{0R} {\rm e}^{ -\kappa_{R} r^{2}}, \ \ \ 
V_{t} = -V_{0t} {\rm e}^{ -\kappa_{t} r^{2}}, \ \ \  
V_{s} = -V_{0s} {\rm e}^{ -\kappa_{s} r^{2}}, \nonumber
\end{eqnarray}
where $P^\sigma $ and $P^r $ are spin- and space-exchange operators, 
respectively. 
An adjustable parameter $u$ determines the strength 
of odd partial-waves between the baryons. 

\vspace*{3mm}
\noindent {(i) $NN$ potential}

The parameters of the Minnesota potential were 
determined so as to reproduce the low-energy $NN$ scattering data. 
This potential is adopted as the $NN$ potential in what follows. 
One nice, noteworthy point 
is that the Minnesota potential reproduces reasonably well 
both the binding energies and sizes of few-nucleon 
systems such as $^2$H, $^3$H, $^3$He and $^4$He\cite{VS95}.  
This is a necessary condition for the present purpose. 
Results of our calculation are insensitive to the 
choice of $u$ because even partial-waves are predominant 
components, so $u$ may be set to unity. 
The Coulomb potential is included in the present calculation. 

\vspace*{3mm}
\noindent {(ii) $\Lambda N$ potential}

Due to the experimental limitation of $\Lambda p$ scattering, 
the $\Lambda N$ interaction is rather poorly known. 
Though various versions of a model description 
for the $\Lambda N$ interaction can reproduce the existing 
$\Lambda p$ scattering data, the properties given by them 
are different from each other. For example, 
the $^1S_0$ and $^3S_1$ phase shifts scatter depending on the model, 
though they are quite important for binding the $\Lambda$-hypernuclei.

The relative importance of $^1S_0$ and $^3S_1$ $\Lambda N$ potentials 
may be estimated in a simple {\it core nucleus}+$\Lambda$ model 
for the $s$-shell $\Lambda$ hypernuclei\cite{DHT72}.  
The $\Lambda$ separation energy, $B_{\Lambda}$, may be estimated by 
$B_{\Lambda}=-(A-1)\langle V_R \rangle - N_s\langle V_s \rangle 
- N_t\langle V_t \rangle $ when the kinetic energy contribution is 
neglected, where $\langle V_s \rangle$, for example, is the average 
of the \LN potential matrix elements in singlet states, 
and the number $N_s (N_t)$ of \LN pairs in the singlet 
(triplet) state is obtained by 
\begin{equation}
\left(
\begin{array}{c}
N_{t}\\
N_{s}\\
\end{array}
\right)
=\left\langle\sum_{i=1}^{A-1}{1\over 2}(1\pm 
P_{i\Lambda}^\sigma)\right\rangle=
{A-1 \over 2} \pm \left({A-1 \over 4}+
\left\langle{\bfi J}_{c}\cdot{\bfi S}_{\Lambda}\right\rangle\right),
\label{EQOFTSWGTS}
\end{equation}
\par\noindent
where ${\bfi J}_{c}$ is the spin of the core nucleus, 
${\bfi S}_{\Lambda}$ is the spin of $\Lambda$, and the total 
angular momentum ${\bfi J}$ of the $s$-shell $\Lambda$-hypernuclei is 
${\bfi J}$=${\bfi J}_{c}$+${\bfi S}_{\Lambda}$. 
Table~\ref{TSWGTS} lists the number of pairs obtained by assuming 
that the spin of the nuclear core remains unchanged 
by the addition of $\Lambda$. 
The fact that the ground state spin of $_\Lambda^3$H is 
$J=\frac{1}{2}$ but not $J=\frac{3}{2}$ indicates that 
the $^1S_0$ potential is more attractive than the $^3S_1$ potential. 
This conclusion is also consistent with the fact 
that the ground states of the $A\!=\!4$ hypernuclei, $_\Lambda^4$H and 
$_\Lambda^4$He, have $J\!=\!0$ instead of $J\!=\!1$. The energy splitting 
of the two states with different $J$ values is expected to give 
us some information on the spin-dependent 
part of the $\Lambda N$ interaction. The above argument is 
qualitative. It is noted that in reality  
the dynamical effect of the $\Lambda N$-$\Sigma N$ coupling modifies 
the contribution of the $^1S_0$ and $^3S_1$-$^3D_1$ $\Lambda N$ 
interaction components from the value of Table~\ref{TSWGTS}, 
as shown in Ref. \cite{Miy95} for the 
$A\!=\!3,\, J\!=\!{1\over 2}$ system. 

We adopt, for the $\Lambda N$ potential, the same form (\ref{MINNPOT}) 
as the $NN$ potential. This ansatz is reasonable because of the 
similarity between $NN$ and $\Lambda N$, which is expected from 
the flavor $SU(3)$ symmetry. In particular the $^1S_0$ $\Lambda N$ 
channel is dominated by $(\lambda \mu)=(22)$ $SU(3)$ symmetry\cite{Nak95}, 
which is exactly the symmetry of the $^1S_0$ $NN$ channel. 
The CSB term of the \LN potential will be considered in subsect. 3.4.

The potential depth and range were determined as follows. First we 
determined the potential parameters so as to reproduce the \LN $S$-wave 
phase shifts predicted by the NF\cite{NF79} 
or the FSS\cite{FSS96} model (see Fig.~\ref{LNPHSFT}). 
Then the binding energies of the $A\!=\!3,\,4$ 
$\Lambda$-hypernuclei were calculated by using this phase equivalent 
potential. By taking a suitable combination of these two different 
potentials we attempted to reproduce the binding energy data. 
The $S$-wave phase shifts calculated by the resultant potential 
was then fitted by a three-range Gaussian potential (\ref{MINNPOT}). 
Of course the last procedure does not lead to a unique parameter set. 
In fact we could find a set of the parameters 
for a given value $V_{0R}$ of the repulsive part. 
Three sets of parameters determined in this way 
are listed in Table~\ref{LNPOTPARA}.

The potentials of sets A and B, used in Refs. \cite{Nem99} and \cite{Suz99}, 
have a moderate height of the central repulsion, while the repulsion of the 
set C potential is chosen to be exceedingly large. The potentials of 
quite different core heights are constructed in order 
to examine the extent to which 
the binding energy and the quark Pauli effect depend on the potential 
shape, that is the dependence of the energy on off-shell properties of the 
potential. Though the potential shapes are different from each other 
as shown in Fig.~\ref{LNPOTSHP}, they give practically the 
same low-energy scattering parameters, i.e., the phase shifts, 
the scattering lengths and the effective ranges. 
The $S$-wave phase shifts calculated by these effective 
potentials are compared in Fig.~\ref{LNPHSFT} to those by 
the NF and FSS models. 
The $^1S_0$ and $^3S_1$ phase shifts predicted by the effective potentials 
grow up to 32 and 19 degrees, respectively. This difference 
in the phase shifts 
is necessary to explain the energy splitting of the $0^+$ and $1^+$ 
states of the $A$=4 systems. The \LN 
potential used by Dalitz {\it et al.}\cite{DHT72} gives similar 
phase shifts between $^1S_0$ and $^3S_1$ states.

The parameter $u$ of the \LN potential is rather insensitive to the 
binding energy data of the $A\!=\!3-5$ hypernuclei because the \LN 
$S$-waves play a predominant role. This parameter was 
determined to be $u\!=\!1.5$ so as to fit the forward and backward ratio 
of the $\Lambda p$ scattering data\cite{LpAlex,LpSech}, as was done by 
Dalitz {\it et al}\cite{DHT72}. The sensitivity to $u$ of the ratio is 
shown in Fig.~\ref{LNFB}.

The total cross section for $\Lambda p$ scattering is 
shown in Fig.~\ref{LPTOT}. 
Our parameter set underestimates the experimental data in low-energy 
region. Some improvement is obtained by introducing the CSB term, 
as we will discuss in subsect.\,3.4. 

Miyagawa {\it et al.}\cite{Miy95} have reported that 
the Nijmegen model NSC89\cite{NSC89} reproduces 
the binding energy of $_\Lambda^3$H. 
According to their analysis, the $^1S_0$ attraction of the \LN interaction 
is quite important for reproducing the binding energy of $_\Lambda^3$H. 
The scattering lengths of the model NSC89 
are $a_s\!=\!-2.78$ fm and $a_t\!=\!-1.41$ fm in a charge symmetric case. 
The corresponding lengths predicted by our effective potentials 
agree with those values within 15 \%; $a_s\!=\!-2.52$ fm 
and $a_t\!=\!-1.20$ fm. 

\vspace*{3mm}
\noindent {(iii) $\Lambda \Lambda$ potential}

Since there are no data for $\Lambda\Lambda$ scatterings, 
$\Lambda\Lambda$ potentials are derived theoretically and 
the soundness of the potential may be tested against the 
binding energy data of the $\Lambda\Lambda$-hypernuclei. 
In the flavor $SU(3)$ symmetry the dominant component of the 
$\Lambda\Lambda$ channel is again $(\lambda \mu)=(22)$\cite{Nak95} but its 
probability is reduced to about 70 \%. 
Only three $\Lambda\Lambda$-hypernuclei have so far been confirmed 
experimentally. The binding energies of these nuclei 
seem to imply that the $\Lambda\Lambda$ interaction is weakly attractive. 
The strength of the $\Lambda\Lambda$ attraction in 
the $\Lambda\Lambda$-hypernuclei may 
be estimated by the $\Delta B_{\Lambda\Lambda}$ value: 
\begin{eqnarray}
\Delta B_{\Lambda\Lambda}(_{\Lambda\Lambda}^{\ \, A}\mbox{X}) 
	&=& B_\Lambda(_{\Lambda\Lambda}^{\ \, A}\mbox{X})
		- B_\Lambda(_{\ \ \ \Lambda}^{A-1}\mbox{X}) \nonumber \\ 
	&=& B_{\Lambda\Lambda}(_{\Lambda\Lambda}^{\ \, A}\mbox{X})
 		- 2B_\Lambda(_{\ \ \ \Lambda}^{A-1}\mbox{X}). 
\label{delbll}
\end{eqnarray}
The experimental $\Delta B_{\Lambda\Lambda}$ values are 
$\Delta B_{\Lambda\Lambda}(_{\Lambda\Lambda}^{\ \ 6}\mbox{He})=4.7\pm0.6$ 
MeV\cite{BLL66}, 
$\Delta B_{\Lambda\Lambda}(_{\Lambda\Lambda}^{\, 10}\mbox{Be})
=4.3\pm0.4$ MeV\cite{BLLBeXLL} and 
$\Delta B_{\Lambda\Lambda}(_{\Lambda\Lambda}^{\, 13}\mbox{B})=4.8\pm0.7$ 
MeV\cite{BLLBXIIILL}. 
(Another interpretation of the last data implies 
$\Delta B_{\Lambda\Lambda}(_{\Lambda\Lambda}^{\, 10}\mbox{B})=-4.9\pm0.7$ 
MeV.) 
We note in passing that the corresponding value for the 
$nn$ attraction may be estimated as 
$$
\Delta B_{nn}=B(^3{\rm H})-2B(^2{\rm H})\cong 4.0 \mbox{MeV},
$$
which is in contradiction with the expectation that the $nn$ 
interaction is more attractive than the \LL interaction. This 
contradiction is probably 
because both $^3{\rm H}$ and $^2{\rm H}$ are rather weakly coupled systems. 

We use two types of $\Lambda\Lambda$ potentials which predict 
weak attraction: One is the potential, denoted OBE-sim\cite{Hiy97}, which is 
phase equivalent to the Nijmegen model D (ND) potential\cite{ND77}, 
and another is the potential, denoted FSS-sim, which is 
phase equivalent to the quark cluster model FSS\cite{Nak99}. 
See Table~\ref{LNPOTPARA} for the potential parameters of FSS-sim. 
The scattering lengths and effective ranges predicted by 
these $\Lambda\Lambda$ potentials are 
\begin{eqnarray*}
a_s &=& -2.80 \mbox{fm}, \ \ \ r_s = 2.81 \mbox{fm} \ \ \ \mbox{(OBE-sim)}\\
a_s &=& -3.01 \mbox{fm}, \ \ \  r_s = 2.16 \mbox{fm} \ \ \ \mbox{(FSS-sim)}\\
a_s &=& -3.01 \mbox{fm}, \ \ \ r_s = 2.14 \mbox{fm} \ \ \ \mbox{(FSS)}.
\end{eqnarray*}
Figure~\ref{LLPHSFTnPOTSHP} compares the $^1S_0$ phase shifts calculated 
by the three models, and exhibits the potential shapes of the OBE-sim and 
FSS-sim models. 

\subsection{Variational Trial Functions}
 
The Hamiltonian $H$ of the system comprising nucleons and 
$\Lambda$ particles is given by 
\begin{equation}
H = T+V = \sum_{i=1}^{A}{{\bfi p}_i^2\over 2m_i}-T_{cm} 
+V^{(NN)}+V^{(\Lambda N)}+V^{(\Lambda \Lambda)}+V^{(C)}, 
\end{equation}
where $V^{(C)}$ is the Coulomb potential between the protons. 
The mass of $N$ is taken as 
$\hbar^2/m_N=41.47$ MeV$\cdot$fm$^2$ and the mass of $\Lambda$ is 
set to be $m_\Lambda/m_N=1.18826$. 
In the present study, we calculate the binding energies of various 
systems up to $A\!=\!6$ in a complete $A$-body treatment. 
The trial function for the eigenfunction of $H$ is given by 
a combination of basis functions:
\begin{eqnarray}
\Psi_{JMTM_T} &=& \sum_{k=1}^{K} 
c_k \varphi_k, 
\label{WFOFGS}
\end{eqnarray}
where the basis function $\varphi_k$ is given by 
\begin{eqnarray}
\varphi_k &=& {\cal A}\{G({\bfi x},A_k)\chi_{kJM}\eta_{TM_T}\}. 
\label{DEFOFWF}
\end{eqnarray}
Here ${\cal A}$ is an antisymmetrizer acting on the 
identical baryons, ${\bfi x}\!=\!({\bfi x}_1, \cdots, {\bfi x}_{A-1})$ 
stands for a set of relative (e.g., Jacobi) coordinates, and 
$\chi_{kSM_S}$ ($\eta_{TM_T}$) is the spin (isospin) function 
of the system. The 
CG, $G({\bfi x}, A_k)$, is defined by 
\begin{eqnarray}
G({\bfi x},A_k) &=& \exp\Big\{-
    \sum_{i<j}^A{\alpha_k}_{ij} ({\bfi r}_i-{\bfi r}_j)^2\Big\} \nonumber \\ 
 &=& \exp\Big\{-\frac{1}{2}
    \sum_{i,j=1}^{A-1}{A_k}_{ij}\,{\bfi x}_i\cdot{\bfi x}_j\Big\}.
\label{DEFOFCORRG}
\end{eqnarray}
The $(A-1)\times (A-1)$ symmetric matrix $A_k$ 
contains $A(A-1)/2$ independent matrix elements which serve as 
nonlinear parameters to characterize the CG basis. 
Since ${\bfi r}_i-{\bfi r}_j$ can be expressed as a combination of 
${\bfi x}_m$, $({\bfi r}_i-{\bfi r}_j)^2$ becomes a combination of 
quadratic terms ${\bfi x}_m\!\cdot\!{\bfi x}_n$, so the two expressions in 
Eq.~(\ref{DEFOFCORRG}) are equivalent. The nonlinear parameters 
${A_k}_{mn}$ can be uniquely expressed in terms of ${\alpha_k}_{ij}$ and 
{\it vice versa}. The parameters ${\alpha_k}_{ij}$ of the 
former expression are more direct 
in representing the range of the correlation between the constituent 
particles\cite{SVM98,CPC97,USU98}. 

The trial function (\ref{WFOFGS}) has clearly the total 
orbital angular momentum $L\!=\!0$. We assume that the hypernuclei 
treated in the present study are well described by only $L\!=\!0$ component.
It is noted, however, that the partial waves corresponding to the coordinate 
${\bfi x}_i$ are not restricted to $l_i$=0, but in general include 
higher orbital angular momenta\cite{SU98,VA98}. 
This is apparent because the basis 
function (\ref{DEFOFCORRG}) 
contains the cross terms ${\bfi x}_i\!\cdot\!{\bfi x}_j$ 
in the exponent of the CG 
and the expansion of those terms into polynomials contains high 
orbital angular momenta. 

The suffix $k$ in $\chi_{kSM_S}$ is used to distinguish possible 
independent spin functions for a given $S$ value\cite{VS95,SVM98}.
For example, in the case where four particles have the total spin $S$=0, 
two independent spin functions are possible. We took into account 
this possibility in the calculation. As for the isospin part 
we used an appropriately coupled function $\eta_{TM_T}$ for a given 
$T$ value.

The set of linear variational parameters ${\bfi c}$=$(c_1,\cdots,c_{K})$ 
of Eq.~(\ref{WFOFGS}) and the energy $E$ are determined 
by Ritz variational principle, which 
leads to the generalized algebraic eigenvalue problem
\begin{equation}
{\cal H} {\bf c} = E {\cal N} {\bf c},
\label{geigen}
\end{equation}
where ${\cal H}$ and ${\cal N}$ are the matrices of the 
Hamiltonian and of the overlap 

\begin{equation}
{\cal H}_{ij}=\langle \varphi_i \vert H \vert \varphi_j \rangle, 
\ \ {\rm and} \ \ \ \
{\cal N}_{ij}=\langle \varphi_i \vert \varphi_j \rangle \ \ \ \  
(i,j=1,\cdots,K).
\end{equation}

The matrix elements of the CGs can be 
calculated analytically for most of important operators. 
This plays a key role in obtaining a precise solution 
for few-body systems. Formulas for the necessary matrix 
elements are given in Refs.~\cite{SVM98,CPC97} and \cite{USU98}. 
See also the appendix where we show how to obtain 
the matrix element for many-particle density operators. 

\subsection{Stochastic Variational Method}

The basis function $\varphi_k$ depends on $A(A-1)/2$ nonlinear 
parameters $\alpha_{k_{ij}}$ or $A_{k_{ij}}$ and also a discrete
parameter to specify the intermediate spin quantum numbers 
in the case when a number of independent spin functions, $g$, is 
$g>1$ for a given $S$ value. These parameters 
define the shape of the basis function and determine how well the 
variational function space contains the true eigenfunction. 
To find the best possible solution, one has to optimize the parameters. 
By assuming that we need a linear
combination of $K$ functions, we face an optimization 
problem of $K(A(A-1)/2+\epsilon)$ 
parameters, where $\epsilon$ is 0 when $g=1$, or 1 otherwise. 
The number of parameters becomes very large 
to get reasonably accurate solutions. For example, in 
the case of \HeVILL, we need $K=200$ roughly and then end up with 3000 
parameters. Moreover,  
Eq.~(\ref{geigen}) must be solved to calculate the energy, 
so that the optimization 
of the parameters is extremely hard in problems of the present type. 

Another serious problem of the minimization of a function 
is the omnipresence of
local minima. A local minimum is the point where the function reaches 
a minimum in a finite interval of variables and the number of such minima 
tends to increase exponentially with the size of the problem.  
The conventional deterministic optimization algorithms (like the Powell or
the conjugate gradient method)\cite{NumRec}  
tend to converge to whichever local 
minimum they first encounter. To avoid such
problems we use the SVM strategy. A key of the SVM is a one by one 
increase of the basis size by searching the best among many random 
trials for the basis function. 

The SVM strategy we have used here 
consists of two procedures, {\it step-by-step} and 
{\it refinement}\cite{VS95,SVM98,VOS97}. 
The first stage is the following trial and error 
procedure to increase the basis dimension:
Let us assume that the sets $A_1,\cdots,A_{k-1}$ and $g_1,\cdots,g_{k-1}$ 
have already been 
selected, and the $(k-1)$-dimensional eigenvalue problem has already been 
solved. Here the index $g_i$ specifies which spin function is  
chosen for $\varphi_i$ from among the $g$ spin functions. 
The next step is the following: 

\vskip 2mm
\par\noindent
{\it step-by-step}
\begin{enumerate}
\item[1.\ \ \,] Different sets $(A_k^1,\cdots,A_k^{n})$ are
generated randomly.

\item[2-1.] For each set $A_k^i$, the $g$ eigenvalue problems of 
$k$-dimension are solved with the $g$ different spin functions 
and the corresponding energies 
$(E_k^{i1},\cdots,E_k^{ig})$ are determined.

\item[2-2.] The parameter of the spin function that produces the lowest 
energy $E_k^i$ among the set $(E_k^{i1},\cdots,E_k^{ig})$ is selected 
to be $g_k^i$ 
for the set $A_k^i$. 

\item[3.\ \ \,] By repeating the above processes 2-1 and 2-2 from $i$=1 to $n$, 
the energies $(E_k^1,\cdots,E_k^{n})$ are determined.

\item[4.\ \ \,] The parameter set $A_k^m$ and $g_k^m$ that produces the 
lowest energy among the set $(E_k^1,\cdots,E_k^{n})$ is selected to 
be the $k$th parameter set for $\varphi_k$.

\item[5.\ \ \,] Increase $k$ to $k+1$ and cycle the processes 1$-$5.
\end{enumerate}
\vskip 2mm
\par\noindent
The essential reason motivating this strategy is the need to sample 
different parameter sets as fast as possible. The advantage 
of this procedure is that it is not necessary to recompute the whole 
Hamiltonian matrix nor is it necessary to perform a new diagonalization 
at each time when a new parameter is generated\cite{SVM98}. This competitive 
selection substantially improves the convergence. 

Obviously the basis size cannot be increased forever.
Moreover, when the $k$th basis state is selected, the previous
states are kept fixed, that is we tried to find the optimal
state with respect to previously selected basis states, but actually
some of the basis states selected earlier might not be so 
important anymore because the succeeding states took over their
roles. So one may include a refining procedure where the previous
states are probed again as described in the following: 

\vskip 2mm
\par\noindent
{\it refinement}
\begin{enumerate}
\item[1.] Random parameter sets $(A_k^1,\cdots,A_k^n)$ are newly generated.

\item[2.] The parameters of the $k$th basis state are
replaced by the new candidates and the energies 
$(E_K^1,\cdots,E_K^{n})$ are calculated.

\item[3.] If the best of the new energies is better than the original
one, then replace the old parameters with the new ones, 
otherwise keep the original ones. 

\item[4.] Cycle this procedure through the basis states from $k=1$ to 
$K$. 
\end{enumerate}
\vskip 2mm
\par\noindent
The selection of the spin function has to be performed in the step 2 
in exactly the same way as in the step-by-step case. 
One may repeat the refining process till no further significant 
improvement is obtained. 
The refinement is useful for suppressing the basis size.

In most of the calculations, we took $n$ as 
$n\sim 50\times{(A(A-1)/2)}$ and generated the matrix $A_k$ 
by changing $\alpha_{k_{ij}}\ (i<j, i=1,\cdots,A-1)$ 
one by one randomly in the interval listed in Table~\ref{PARAM}, where,  
by expressing $\alpha$ as 
\begin{equation}
\alpha={1\over d^2},
\end{equation}
$d$ was chosen in the range of [0, $d_{max}$]. Also 
listed in the table are values of $g$ used in the present calculation.   
We show the convergence of the energy for $^{\ \ 5}_{\Lambda\Lambda}$He 
in Fig.~\ref{h5llconv}. Both procedures of SVM are found to be very 
effective for obtaining solutions with high quality.

A virial ratio $\eta$ is often used to test the accuracy of the 
solution $\Psi$.  
According to the virial theorem, the ratio 
\begin{equation}
\eta = \left|\frac{\langle\Psi|W 
|\Psi\rangle}{2\langle\Psi|T|\Psi\rangle}-1\right|, \ \ \ {\rm with}\ \ \ 
W = \sum_{i=1}^{A}{\bfi r}_i\cdot{\partial V\over \partial{\bfi r}_i}
\label{VIRIAL}
\end{equation}
must vanish for the eigenstate $\Psi$ of the Hamiltonian\cite{SVM98}. 
We will calculate $\eta$ to check the accuracy of our variational 
solution. Note that a spherical symmetric potential $V(|{\bfi r}_i-
{\bfi r}_j|)$ is changed into $rV^{\prime}(r)$ with 
$r=|{\bfi r}_i-{\bfi r}_j|$ by the operation of 
$\sum_{i=1}^{A}{\bfi r}_i\cdot{\partial \over \partial{\bfi r}_i}$. 

\section{Results}

First we show in Table~\ref{rmstab} the root-mean square (rms) distances 
of the particles in the $\Lambda$- and $\Lambda\Lambda$-hypernuclei 
together with 
the typical dimension size $K$ and the virial ratio $\eta$. 
Here the rms distances are defined as follows: 

\begin{equation}
\langle r^2\rangle 
= {1\over A}\Big\langle\Psi\left|
\sum_{i=1}^{A}({\bfi r}_i-{\bfi x}_{A})^2\right|\Psi\Big\rangle,\ \ \ \ \ 
\langle r_{ij}^2\rangle 
= \big\langle\Psi\left|({\bfi r}_i-{\bfi r}_{j})^2\right|\Psi\big\rangle,
\end{equation}
where ${\bfi x}_A$ is the center-of-mass coordinate of the system. 
We see that both $^3_{\Lambda}$H and $^{\ \ 4}_{\Lambda\Lambda}$H, 
among others, have very large rms values, reflecting the very 
small binding energies, as shown below. The rms distance of 
$NN$ decreases by adding more $\Lambda$ particles: By adding  
$\Lambda$ to $^2$H, the 
$\sqrt{\langle r_{NN}^2\rangle}$ value changes from 
3.9 to 3.6 ($^3_{\Lambda}$H) and 3.3 fm ($^{\ \ 4}_{\Lambda\Lambda}$H), 
respectively. A similar but slightly mild change is seen in 
$^3$H and $^3$He as well. In the case of $^4$He, however, we see 
that the rms distance of $NN$ hardly changes by the addition of $\Lambda$.  
Generally speaking, the basis size $K$ needed for an accurate 
solution increases with an increasing number of 
particles and also with an increasing spatial extension of the system. 
Comparing the solutions for the different potentials, 
the potential with an exceedingly large repulsion like the set C or 
the OBE-sim potential makes an accurate solution more challenging. 
It is worthwhile noting that the increase of the basis size 
is rather gentle on these conditions, so that the 
solutions have been obtained in the 
order of $\eta \le 10^{-4}$ in most cases.

\subsection{$\Lambda$-Hypernuclei}

Table~\ref{BEOFSL} shows the results of calculation 
for the $\Lambda$-hypernuclei. 
Since the parameters in Table~\ref{LNPOTPARA} are determined to
fit the binding energies of the $A$=3, 4 $\Lambda$-hypernuclei, 
it is apparent that 
the energies of those nuclei are in good agreement 
with the experimental data including the excited states 
of the $A$=4 system. As noted in subsect.\,2.1, we expect that 
our effective potential has very reasonable relative strength between 
$^1S_0$ and $^3S_1$ states. 

Though the potentials of set A and set C are quite different 
in the shape, they are constructed to be phase equivalent, and 
the binding energies predicted by these potentials are almost 
the same. It is an 
agreeable feature that the phase equivalent potentials predict the 
same energies for all the nuclei. This is a strong point of the 
present calculation with high precision. Of course 
the solution for the case of set C potential requires 
a larger basis dimension than the one for the case of set A potential, 
so that the variational solution must be obtained to an accuracy 
of a few tens of keV regardless of the characteristics of potentials.  
We note that the above nice feature is not preserved in a 
naive calculation such as a {\it frozen core nucleus}+$\Lambda$  
model and that you may draw an erroneous conclusion based on such 
calculations. 
For example, if we describe $_{\Lambda}^5$He 
as an $\alpha+\Lambda$ model where the $\alpha$-particle 
wave function is given by a fixed $(0s)^4$ Slater determinant 
with the size parameter of 1.39 fm, we obtain quite different 
$B_\Lambda(^5_\Lambda\mbox{He})$ values depending on the core height 
of the \LN potential: 3.95 and 2.29 MeV for set A and set B, respectively 
(no anomaly in $B_\Lambda(^5_\Lambda\mbox{He})$!). 
Set C potential does not even bind $_{\Lambda}^5$He. (Here 
$B_\Lambda(^5_\Lambda\mbox{He})$ is defined by the difference 
of the energy of the model $\alpha$-particle ($-$23.94 MeV) and the 
total energy of the $\alpha+\Lambda$ model calculation.) 

As seen in Fig.~\ref{LNPHSFT}, our effective potential is more attractive 
in $^1S_0$ than in $^3S_1$. We asked a question of whether this 
potential is strong enough to bind an excited state of $_\Lambda^3$H with 
$J^\pi=\frac{3}{2}^+$. The calculation showed that the energy 
of the system approaches the deuteron energy, that is, the 
potential accommodates no such a bound state. 

It has been known for a long time that the CSB effect of the 
\LN interaction 
is manifestly exhibited in the $B_{\Lambda}$ values of 
the $A$=4 hypernuclei. The data suggest that the $\Lambda p$ 
interaction is stronger than the $\Lambda n$ interaction. 
Though the $\Lambda N$-$\Sigma N$ coupling or $\Lambda$-$\Sigma^0$ 
mixing is expected to play the significant role in the CSB, 
no conclusive understanding of the mechanism of CSB has been obtained yet. 
Both potentials of sets A and C are charge symmetric.  
An extension to a calculation including the CSB effect 
will be discussed later. 

The result for $_\Lambda^5$He confirms the long standing 
problem, i.e., the potential fitted to the $A$=3, 4 data 
overbinds $_\Lambda^5$He by about 2 MeV. 
We will come back to this anomaly in sect.\,4, where the 
effects of the quark Pauli principle are discussed.

\subsection{$\Lambda\Lambda$-Hypernuclei}

Table~\ref{BEOFDL} shows the results of the 
$\Lambda\Lambda$-hypernuclei. 
The energies calculated with different \LN potentials are almost 
the same. 
We have examined if a lightest system of $A\!=\!3$ forms a 
bound state, 
but have not found a stable $N\!\Lambda\Lambda$ system with the 
$\Lambda\Lambda$ potential either OBE-sim or FSS-sim. 
We have then asked a question of 
whether or not a bound state is formed for $A\ge 4$ system.  
We have calculated the energies for various spin $S$ and isospin $T$ 
states: 
$(S,T)=(0,0),\, (0,1),\, (1,0)$ and $(1,1)$ for $_{\Lambda\Lambda}^{\ \ 4}$H, 
and $(S,T)=(0,1)$ and $(1,1)$ for $_{\Lambda\Lambda}^{\ \ 4}$He 
or $_{\Lambda\Lambda}^{\ \ 4}$n. 
We find that a lightest bound state of $\Lambda\Lambda$-hypernuclei 
is only $_{\Lambda\Lambda}^{\ \ 4}$H 
with $(S,T)=(1,0)$ in consistency with the conclusion 
of Ref.~\cite{Nak90}. See Table~\ref{BEOFDL}. 
The binding energy is so small that its existence may be strongly 
subjected to the strength of the $\Lambda\Lambda$ potential. 
This state has a structure of a deuteron plus $\Lambda\Lambda$ that is 
coupled to $S$=0. Results for $A$=5 $\Lambda\Lambda$-hypernuclei 
are also listed in Table~\ref{BEOFDL}. It is very likely that 
both $_{\Lambda\Lambda}^{\ \ 5}$H and $_{\Lambda\Lambda}^{\ \ 5}$He are 
bound. Experimental 
search for these $\Lambda\Lambda$-hypernuclei and measurements 
of their binding energies will provide us with valuable information on 
the $\Lambda\Lambda$ interaction.  

The main component of $\Lambda\Lambda$ is the $^1S_0$ state 
for the $A$=4$-$6 $\Lambda\Lambda$-hypernuclei, so 
the total angular momentum $J$ of each hypernucleus is equal to 
that of the core nucleus. The $J$ and 
$T$ values are $(J,T)=({1\over 2},{1\over 2})$ 
for $_{\Lambda\Lambda}^{\ \ 5}$H and $_{\Lambda\Lambda}^{\ \ 5}$He, 
and (0,0) for $_{\Lambda\Lambda}^{\ \ 6}$He. 
The contribution of the $^3S_1$ and the $^1S_0$ components 
of the \LN potential to the potential energy is determined by $N_t$ and 
$N_s$ values. Using an arithmetic similar to Eq.~(\ref{EQOFTSWGTS}) 
we obtain $N_{t}$=${3\over 2}(A-2)$ and $N_{s}$=${1\over 2}(A-2)$. 
Their ratio is 
$N_t:N_s=3:1$ for all these $\Lambda\Lambda$-hypernuclei, as in 
the case of \HeVL. 

The energy of $B_{\Lambda\Lambda}(_{\Lambda\Lambda}^{\ \ 6}\mbox{He})$ 
is again overbound compared to the empirical value, just as in 
the case of $B_\Lambda(_\Lambda^5\mbox{He})$ shown in Table~\ref{BEOFSL}.  
The calculated value 
of $\Delta B_{\Lambda\Lambda}(_{\Lambda\Lambda}^{\ \ 6}\mbox{He})$ 
is, however, in reasonable agreement with the experimental 
value of $4.7\pm0.6$ MeV: $\Delta B_{\Lambda\Lambda}$ = 4.3 MeV 
for OBE-sim $\Lambda\Lambda$ potential 
and 5.2 MeV for FSS-sim $\Lambda\Lambda$ potential. We thus expect that 
the strength of the $\Lambda\Lambda$ potentials employed in the 
present study is nearly right. Rather we guess that the same anomaly 
known in $_\Lambda^5$He appears in $_{\Lambda\Lambda}^{\ \ 6}$He as well.

\subsection{Density Distributions and Correlation Functions}

Obviously the density distribution and the correlation function between 
the particles provide us with more profound information on the 
structure of the system than just the average distances of the particles. 
It is also of particular interest to know 
the extent to which the core nucleus is distorted by adding the $\Lambda$ 
particles. For this purpose we define 
the distribution function of $N$ or $\Lambda$ by 

\begin{equation}
\rho({\bfi r})=\langle\Psi\left|
\delta({\bfi r}_{i}-{\bfi R}_c-{\bfi r})\right|\Psi\rangle, 
\label{RHOFN}
\end{equation}
where ${\bfi r}_i$ denotes the position vector of $N\mbox{ or }\Lambda$, 
and ${\bfi R}_c$ is the center-of-mass coordinate of the core nucleus.  
Therefore ${\bfi r}$ is a position vector from the core nucleus.  
The density defined above is different from 
the conventional one which is calculated as a function of a 
position vector measured from 
the center-of-mass coordinate of the system. The $N$ or $\Lambda$ density 
defined here seems more direct to represent 
the change with the addition of $\Lambda$. See the appendix for the 
method of calculation of the density and the correlation function etc. 

Figures~\ref{h4ll} $\!-\!$~\ref{h6ll} show the density distributions of 
$N$ and $\Lambda$ 
when $\Lambda$ particles are added to the core nucleus 
of $^2$H, $^3$H or $^4$He, respectively. 
The set A \LN and OBE-sim \LL potentials are used but  
other choices of the potentials do not change the result very much. 
As seen from these figures, the density distributions of 
$N$ and $\Lambda$ are quite different. 
The $\Lambda$ distribution reaches further more outwardly than the 
$N$ distribution. The systems of $_\Lambda^3$H and 
$_{\Lambda\Lambda}^{\ \ 4}$H are extremely spread out because the 
deuteron is weakly bound and in addition these hypernuclei have very 
small separation energies. The maximum probability of 
finding $\Lambda$ occurs at or near the point where the nucleon 
probability falls to half of its maximum probability. 
Figure~\ref{h5ll} shows that the $\Lambda$ distribution is  
considerably different between the ground and excited states 
of $_\Lambda^4$H 
and that the core distortion is larger in the ground state 
than in the excited state.  
We see that the addition of $\Lambda$ particles makes the core 
nucleus shrink and accordingly 
the $\Lambda$ density move to the inner region. 

The function $C({\bfi r})$ defined by 

\begin{equation}
C({\bfi r})=\langle\Psi\left|\delta({\bfi r}_{i}-{\bfi r}_j-{\bfi r})
\right|\Psi\rangle 
\label{CRFN}
\end{equation}
gives the information on the correlation of the constituent particles. 
The correlation function of \LL is calculated with the 
set A \LN and the OBE-sim \LL potentials and compared in Fig.~\ref{cfll} 
for the $\Lambda\Lambda$-hypernuclei, $_{\Lambda\Lambda}^{\ \ 4}$H, 
$_{\Lambda\Lambda}^{\ \ 5}$H and $_{\Lambda\Lambda}^{\ \ 6}$He.
It is noted that the correlation function has vanishingly small 
amplitudes at short 
distances, as it should because the OBE-sim \LL potential has 
a strong central repulsion. Since the CGs 
have large amplitudes at short distances of the particles, this 
vanishing of the correlation function indicates that the 
CG basis is sufficiently flexible to reproduce such characteristics 
and that the SVM has been practical in selecting suitable basis sets. 
The behavior of the correlation function reflects the 
$\Delta B_{\Lambda\Lambda}$ value defined by Eq.~(\ref{delbll}): 
The calculated 
$\Delta B_{\Lambda\Lambda}$ values are 0.06, 1.2 and 4.3 MeV 
for $_{\Lambda\Lambda}^{\ \ 4}$H, 
$_{\Lambda\Lambda}^{\ \ 5}$H and $_{\Lambda\Lambda}^{\ \ 6}$He, 
respectively. The wider the spatial extension of the correlation 
function, the weaker the $\Lambda\Lambda$ interaction. 
We see that the two $\Lambda$ particles, particularly 
in $_{\Lambda\Lambda}^{\ \ 4}$H, 
spend almost all the time outside their interaction range. 
The existence of $_{\Lambda\Lambda}^{\ \ 4}$H crucially depends 
on the strength of the \LL potential. 
The peak position of the correlation function moves depending on the 
size of the underlying core nucleus. 

To visualize the structure of the system, it is useful to 
calculate the two-particle (particularly \LL of the 
${\Lambda\Lambda}$-hypernuclei) 
distribution function given by 

\begin{equation}
D({\bfi r}, {\bfi r}^{\prime})=\langle\Psi\left|
\delta({\bfi r}_i-{\bfi R}_c-{\bfi r})
\delta({\bfi r}_j-{\bfi R}_c-{\bfi r}^{\prime})
\right|\Psi\rangle. 
\label{TPDFN}
\end{equation}
This function depends on the three quantities, 
$r,\, r^{\prime}$ and $\theta$, in the case of 
$L=0$ wave function, 
where $\theta$ is the angle between ${\bfi r}$ and ${\bfi r}^{\prime}$; 
${\bfi r}\!\cdot\!{\bfi r}^{\prime}=r r^{\prime}{\rm cos}\,\theta$.  
In Fig.~\ref{tpden}, the two-$\Lambda$ distribution function 
$D({\bfi r}, {\bfi r}^{\prime})$ weighted by 
$r^2{r^{\prime}}^2{\rm sin}\theta$ is displayed for \HVLL. 
To draw this 
function we first searched for a point where the probability, 
$r^2{r^{\prime}}^2{\rm sin}\,\theta \,D({\bfi r}, {\bfi r}^{\prime})$, 
reaches a maximum. The maximum for $^{\ \ 5}_{\Lambda\Lambda}$H 
appears at the point 
of $r=r^{\prime} $=1.8 fm and $\theta$=47$^{\circ}$. Drawn in 
Fig.~\ref{tpden} is the distribution of $\Lambda$ when the 
other $\Lambda$ was placed at that point. Though the point determined 
above may suggest a configuration of an isosceles triangle 
with side lengths of 1.8, 1.8 and 1.4 fm, the distribution is actually 
spread out in a wide region, so that the geometry is to be understood 
rather loose. For example, the rms distance of \LL is 
as large as 3.5 fm, much larger than 1.4 fm.
 
The core nucleus can be distorted by the addition of $\Lambda$ particles, 
that is, the nucleus never remains in its ground state in the 
hypernucleus but may have mixing in of other states. 
To calculate the probability that the core nucleus remains 
in its ground state, we define the spectroscopic factor $S$ by 

\begin{equation}
S= \sum_{M_c, M_s}\int d{\bfi r} \left\{g_{M_c M_s}({\bfi r})\right\}^2,
\end{equation}
where the spectroscopic amplitude $g_{M_c M_s}({\bfi r})$ 
for the $\Lambda$-hypernuclei is given by

\begin{equation}
g_{M_c M_s}({\bfi r})=\big\langle
\Psi_{J_cM_cTM_T}(^{A-1}{\rm X})\chi_{{1\over 2}M_s}(\Lambda) 
\delta({\bfi x}_{A-1}-{\bfi r}) | 
\Psi_{JMTM_T}(_\Lambda^A{\rm X})\big\rangle.
\end{equation}
Here ${\bfi x}_{A-1}={\bfi r}_{A}-{\bfi R}_c$ is the 
distance vector of $\Lambda$ from 
the center-of-mass of the core nucleus $^{A-1}$X, 
and $\Psi_{J_cM_cTM_T}(^{A-1}{\rm X})$ 
is the ground state wave function of the core nucleus. 
If no distortion of the core nucleus is present in the hypernucleus, 
$\sum_{M_c, M_s}\left\{g_{M_c M_s}({\bfi r})\right\}^2$ 
reduces to the density distribution of $\Lambda$ that is defined by 
Eq.~(\ref{RHOFN}). 

The spectroscopic factor for the $\Lambda\Lambda$-hypernucleus 
is calculated in a similar way. The spectroscopic amplitude 
may be defined by 

\begin{equation}
 g_{M_c}({\bfi r}, {\bfi r}^{\prime})
= \big\langle
\Psi_{J_cM_cTM_T}(^{A-2}{\rm X})\chi_{00}(\Lambda\Lambda) 
\delta({\bfi x}_{A-2}-{\bfi r})
\delta({\bfi x}_{A-1}-{\bfi r}^{\prime}) | 
\Psi_{JMTM_T}(_{\Lambda\Lambda}^{\ \ A}{\rm X})\big\rangle, 
\end{equation}
where ${\bfi x}_{A-2}={\bfi r}_{A-1}-{\bfi R}_c$ and 
${\bfi x}_{A-1}={\bfi r}_{A}-{\bfi R}_{c+\Lambda}$ with 
${\bfi R}_{c+\Lambda}$ denoting the center-of-mass of 
the core nucleus and the $\Lambda$. Note that the two $\Lambda$ 
particles were assumed to be in spin singlet. The spectroscopic 
factor $S$ is then obtained by  
$\left\{g_{M_c}({\bfi r}, {\bfi r}^{\prime})\right\}^2$ summed 
over $M_c$ and integrated over ${\bfi r}$ and ${\bfi r}^{\prime}$. 

Table~\ref{SPFACTOR} lists the spectroscopic factors 
calculated for $\Lambda$- and $\Lambda\Lambda$-hypernuclei. 
We confirm, in conformity with the results of 
Figs.~\ref{h4ll} $\!-\!$~\ref{h6ll}, that adding more $\Lambda$ 
particles results in smaller $S$ values, that is, it leads to the 
larger distortion of the core nucleus, as expected. 
There is no experimental information 
on $S$ values but just one theoretical 
value is available for $^3_{\Lambda}$H. According to Ref.~\cite{Miy95} 
the value is 0.987, which is in very reasonable agreement 
with the present estimate (0.991)
in spite of the use of quite different potentials.  
The $S$ values 
of $^5_{\Lambda}$He and $^{\ \ 6}_{\Lambda\Lambda}$He are fairly 
large compared to those of the other nuclei, which may indicate 
that the $^4$He core is considerably stable. 
As already pointed out, however, these results do not necessarily 
substantiate that the frozen core model is a perfect model 
even for the light $s$-shell hypernuclei. A small distortion 
becomes sometimes crucial for the accurate evaluation of the 
binding energies.

\subsection{Charge Symmetry Breaking}

The experimental data of the $A$=4 system imply the  
CSB of the $\Lambda N$ interaction. 
The CSB is observed for both the ground and excited states:
\begin{eqnarray}
\Delta B_\Lambda &=& B_\Lambda(_\Lambda^4\mbox{He}) 
   - B_\Lambda(_\Lambda^4\mbox{H}) = 0.35\pm 0.07 \mbox{MeV}, \nonumber \\
\Delta B_\Lambda^{\, *} &=& B_\Lambda(_\Lambda^4\mbox{He}^*) 
   - B_\Lambda(_\Lambda^4\mbox{H}^*) = 0.24\pm 0.15 \mbox{MeV}. 
\label{CSBDATA}
\end{eqnarray}
Bodmer and Usmani\cite{Bod85} investigated the CSB phenomenologically. 
They suggest that the CSB interaction is effectively spin independent, 
that is, it has no ${\bfi \sigma}_\Lambda\!\cdot\!{\bfi \sigma}_N$ 
dependence. 
In their argument, however, the ground and excited states were not 
treated simultaneously. 
They modified the strength of the $\Lambda N$ attraction 
with respect to the ground or excited state, 
in spite of the fact that the strength 
of ${\bfi \sigma}_\Lambda\!\cdot\!{\bfi \sigma}_N$ interaction 
is different from each other in the case of $J^\pi=0^+$ and $1^+$ states. 
Dalitz {\it et al.}\cite{DHT72} studied the CSB effect for both cases of the 
spin-dependent and spin-independent CSB potentials. 
However, they focused only on the CSB effect of the ground states 
of $_\Lambda^4$H and $_\Lambda^4$He. 
Since the accuracy of the present calculation is high enough to discuss 
the CSB, we attempted to 
determine phenomenologically a $\Lambda N$ interaction which 
includes the CSB effect.

We introduce the CSB term in both components of the triplet and singlet 
$\Lambda N$ potentials (\ref{MINNPOT}) as follows 
\begin{equation}
V_{0t} \rightarrow V_{0t} - v_t\tau_3, \ \ \ 
V_{0s} \rightarrow V_{0s} - v_s\tau_3,
\label{csbpot}
\end{equation}
where $\tau_3$ is the third component of the isospin matrices:   
\begin{equation}
\tau_3 |n\rangle = + |n\rangle, \ \ \ 
\tau_3 |p\rangle = - |p\rangle. 
\end{equation}
The modification (\ref{csbpot}) of the strength  
implies the CSB potential of the form 
\begin{equation}
V_{\rm CSB}= \left\{{1 \over 4}(3v_t+v_s)\tau_3 
+{1 \over 4}(v_t-v_s)
({\bfi \sigma}_\Lambda\!\cdot\!{\bfi \sigma}_N)\tau_3\right\}
{\rm e}^{-\kappa r^2}.
\end{equation}
(Here $\kappa_t$ and $\kappa_s$ are assumed to be the same to simplify 
the expression but no such an assumption was made in the calculation.) 
If $v_t$ and $v_s$ have the same sign, the spin dependence of the CSB 
potential becomes weak. 

Starting from the set A \LN potential in Table~\ref{LNPOTPARA}, we 
determined those $v_t$ and $v_s$ values of the CSB potential which 
reproduce the data as suggested in Eq.~(\ref{CSBDATA}). For this purpose 
we examined the energy dependence of the $A$=4 hypernuclei on the set of 
$v_t$ and $v_s$ values around $(v_t, v_s)$=(0, 0). After repeating 
calculations for different sets of parameters, we found a set of 
values which reproduces both 
$\Delta B_\Lambda$ and $\Delta B_\Lambda^{\, \ast}$ very well: 
\begin{equation}
v_t = 3.30\ \mbox{MeV}\ \ \ {\rm and} \ \ \ 
v_s = 2.65\ \mbox{MeV}. 
\label{CSBPARA}
\end{equation}
Both $v_t$ and $v_s$ values are positive in sign, so that the CSB 
potential determined above is weakly spin-dependent 
in accordance with the conclusion of Ref.~\cite{Bod85}. 
With this CSB potential the $\Lambda p$ phase shifts in $^1S_0$ and $^3S_1$ 
states increase to 34 and 21 degrees, respectively. 

As listed in Table~\ref{BEOFSL}, the calculated binding 
energies with this CSB potential are 
in good agreement with experimental data.  
The change of 
$B_{\Lambda\Lambda}$ with the CSB potential is 
listed in Table~\ref{BEOFDL}. The CSB does not alter the prediction for 
possible existence of $_{\Lambda\Lambda}^{\ \ 5}$H and 
$_{\Lambda\Lambda}^{\ \ 5}$He. 

The CSB potential produces a very minor change in energy for those 
hypernuclei 
which include the same number of protons and neutrons such as 
$_\Lambda^3$H, $_{\Lambda\Lambda}^{\ \ 4}$H, $_\Lambda^5$He 
and $_{\Lambda\Lambda}^{\ \ 6}$He. 
 
The scattering parameters calculated with the CSB potential 
(\ref{CSBPARA}) are compared in Table~\ref{SCTLNGOFCSB} 
to predictions by different models. 
Some models by e.g. Nijmegen group\cite{NF79,NSC89,NSC97,Nij73} 
give a spin-dependent CSB interaction and predict 
the strengths of the $\Lambda N$ attraction in the order of 

\begin{equation}
|a_s^{\Lambda n}| > |a_s^{\Lambda p}| > |a_t^{\Lambda p}| > |a_t^{\Lambda n}|.
\label{ORDOFCSB}
\end{equation}
On the contrary, our potential predicts the opposite order 
in the $^1S_0$ part: As expected, the $\Lambda p$ interaction 
becomes more attractive than the $\Lambda n$ interaction. 
In the case of $^3S_1$ both the Nijmegen and our phenomenological potentials 
predict that the $\Lambda p$ interaction is 
more attractive than the $\Lambda n$ interaction.
It is noted that the introduction of the CSB determined in this study 
improves a fit to the $\Lambda p$ total cross section as displayed in 
Fig.~\ref{LPTOT}.

\section{Quark Pauli Effect}

\subsection{Pauli-forbidden Configuration}
As shown in the previous section, the calculated $\Lambda$ 
separation energy of $_\Lambda^{\, 5}$He is overbound by 
about 2 MeV, compared to experiment. 
The $\Lambda\Lambda$ separation energy 
of $_{\Lambda\Lambda}^{\ \ 6}$He is found to be also largely 
overbound. To resolve 
these discrepancies is one of the most challenging problems in 
$s$-shell hypernuclei. Shinmura {\it et al.}\cite{Shi84} argued that 
the noncentral components of the \LN and $NN$ potentials 
plays an important role. 
The effect of the $\Lambda$-$\Sigma$ conversion may become important, 
as shown by Miyagawa {\it et al.}\cite{Miy95}, through 
the $^3S$-$^3D$ tensor coupling in the $\Lambda N$-$\Sigma N$ channel. 
An {\it ab initio} calculation using realistic potentials 
has not been performed until now because of the complexity of the 
potentials. Even if it has been done 
in some way, there is so large uncertainty in the realistic potentials 
that such a calculation may not give a perfect answer. 
The premise we are taking in this study is that the effects of the 
noncentral forces and the $\Lambda$-$\Sigma$ conversion etc. are 
renormalized in the effective potential which is determined so as to 
fit the binding energies of the $A$=3, 4 $\Lambda$-hypernuclei. 
We thus pay attention to other mechanism, a quark Pauli effect, 
for examining this problem in this section. 
The quark Pauli effect was considered in Ref.\cite{Hun84} on the basis 
of the quark shell model where the quarks were allowed to move in the 
whole hypernucleus. 
In contrast to this approach, we treat the Pauli effect by confining 
three quarks inside the baryons. 
Parts of the results 
along this line have been reported in Refs.~\cite{Nem99} and \cite{Suz99}. 

So far we have treated the $\Lambda$ particle as distinguishable from 
the nucleon $N$.  
Therefore all the baryons in $_\Lambda^5$He or 
$_{\Lambda\Lambda}^{\ \ 6}$He, for example, may 
occupy a small region at the same time if that configuration is favorable 
to the energy gain. 
In the quark model of baryons, however, $\Lambda$ and $N$ share 
quarks of the same flavor, and then in a $\Lambda$-hypernuclei with $A\ge 5$ 
there may be an increasing chance that more than six $u$ or $d$ quarks occupy 
the same orbit as the baryons come close to each other. Of course 
this chance is forbidden from the quark Pauli principle, so that it 
should be excluded in a calculation\cite{Nem99,Suz99}. 
This type of quark Pauli principle should act even in normal nuclei but its 
effect is taken into account by the Pauli principle for the $N$. 
There is no way, however, to take into account the quark Pauli effect in those 
calculations which do not consider the quark substructure of the baryon. 
The aim here is to estimate how much the $\Lambda$ separation energy 
changes by the mechanism arising 
from the quark Pauli effect. We perform this estimation 
by using only those wave functions which are still expressed in 
terms of the baryon coordinates. The effect to be discussed here is concerned 
with a specific forbidden configuration, which is a special effect 
of a more general antisymmetrization effect of the quarks\cite{Tak86}. 

A specific quark Pauli-forbidden configuration comprising five baryons was 
considered in Refs.~\cite{Nem99} and \cite{Suz99}: 

\begin{equation}
\psi_{\rm F}({\bfi x})=\Big({\pi b^2 \over 3}\Big)^{-{3\over 4}}
 {\exp}\left\{-{3 \over 2b^2}\sum_{i=1}^5
({\bfi r}_i-{\bfi R}_5)^2\right\}, 
\label{WFOFQP}
\end{equation}
where ${\bfi R}_5$ is the center-of-mass coordinate 
of the five baryons. 
This configuration was derived by assuming that a baryon is a 
three-quark system and that the orbital motion of each quark is 
described by a $(0s)$ harmonic-oscillator 
function, $\exp\{-{\bfi r}^2/2b^2\}$. The parameter $b$ determines 
the size of the baryon. If the simple $(0s)^3$ wave function ought 
to reproduce the charge radius of proton (0.86 fm)\cite{Sim80}, then 
$b$ becomes 0.86 fm. A more reasonable value of $b$ would be 
slightly smaller than 0.86 fm if other effects such as 
meson clouds etc. are considered in the charge radius. 
We will examine the binding energy change in the range of 
$b$=0.6$-$0.86 fm.
 
Now the trial function must be 
subjected to remain in the subspace which is orthogonal to the forbidden 
state, that is, $\langle\psi_{\rm F}|\Psi\rangle=0$. 
We thus modify Eq.~(\ref{WFOFGS}) by  
\begin{eqnarray}
\Psi_{JMTM_T} &=& \sum_{k=1}^{K} 
c_k\left(\varphi_k-\sum_a\Gamma_a\langle\Gamma_a|\varphi_k\rangle\right), 
\label{WFOFGSQP} 
\end{eqnarray}
where $\Gamma_a$ is an orthonormal set which spans the Pauli-forbidden 
space. In the case of $A$=5, $\Gamma_a$ can simply be represented by 
${\cal A}\{\psi_{\rm F}\chi\eta\}$. 
For $A\ge 6$, however, the construction of $\Gamma_a$ is never trivial. 
The system $^{\ \ 6}_{\Lambda \Lambda}$He, for example, 
has three types of forbidden states (\ref{WFOFQP}) 
comprising $ppnn \Lambda$, $ppn \Lambda \Lambda$, or 
$pnn \Lambda \Lambda$ baryons. 
A method proposed in Ref.~\cite{Suz99} enables us to eliminate 
practically the Pauli-forbidden components. 
The reader is referred to Ref.~\cite{Suz99} for its detail. 
The basis functions $\varphi_k$ were newly selected by the SVM. 
The accuracy of the solution is an essential ingredient 
in this calculation because the solution must be accurate enough to be 
sensitive to small components produced by the quark Pauli principle.

\subsection{Hypernuclei with $A$=5 and 6}

Table~\ref{SEPENEQP} shows the dependence of 
the $\Lambda$ separation energy on the baryon size 
for the $A$=5, 6 hypernuclei. The case of $b$=0 is no quark Pauli 
calculation. The energy change is insignificant for $b \le 0.6$ fm. 
As expected, a significant binding energy reduction is found to 
occur at $b=0.86$ fm in $^{5}_{\Lambda}$He and 
$^{\ \ 6}_{\Lambda\Lambda}$He, while the reduction is much less 
in $^{\ \ 5}_{\Lambda\Lambda}$H and $^{\ \ 5}_{\Lambda\Lambda}$He. 
The energy of $^{\ \ 6}_{\Lambda \Lambda}$He is apparently 
overbound in no quark Pauli calculation. The elimination of the 
forbidden states 
produces a surprisingly large reduction in the binding energy, 
the order of half a MeV even at $b$=0.6 fm, which  
is very favorable to the anomaly problem. These 
results can be understood from the fact that 
$^{\ \ 5}_{\Lambda \Lambda}$H 
and $^{\ \ 5}_{\Lambda \Lambda}$He are spatially more extended 
than $^{5}_{\Lambda}$He and $^{\ \ 6}_{\Lambda\Lambda}$He. 
(See Table~\ref{rmstab} and Figs.~\ref{h4ll} $\!-\!$~\ref{h6ll}.) 
Since the $b$-dependence of the quark Pauli effect is very moderate 
in $_{\Lambda\Lambda}^{\ \ 5}$H and $_{\Lambda\Lambda}^{\ \ 5}$He, 
a measurement of their binding energies will be quite useful for 
the determination of the \LL attraction. 

As the forbidden state (\ref{WFOFQP}) is a spatially compact configuration, 
one may conceive that the energy change be dependent on the 
characteristics of potentials, especially the height of the central 
repulsion. Table~\ref{SEPENEQP2} compares $B_{\Lambda}$ and 
$B_{\Lambda\Lambda}$ values for different sets of the potentials. 
The variation of the $\Lambda$ separation energy for the different 
potentials is very moderate, so we may conclude that the results of 
these tables indicate the general trend of the quark Pauli effect in light 
$\Lambda$- and $\Lambda\Lambda$-hypernuclei. 

To explain the mild dependence on the potentials we derive an 
approximate expression for the energy change as follows. 
Let $\Psi_0$ and $E_0$ denote the ground state wave function 
and the energy of no quark Pauli calculation. 
When the quark Pauli effect is taken into account, 
the ground state wave function changes and, since the change is 
expected to be small, the wave function may be approximated as 
\begin{equation}
\Psi ={1 \over \sqrt{1-\varepsilon^2}}\left(\Psi_0 - 
\varepsilon\psi_{\rm F} \right)\ \ \ \ 
{\rm with}\ \ \ \ \varepsilon = \langle\psi_{\rm F}|\Psi_0\rangle.
\end{equation}
The energy change is given by 
\begin{equation}
E - E_0 = \langle\Psi|H|\Psi\rangle - E_0 
	\approx \varepsilon^2 \langle\psi_{\rm F}|H|\psi_{\rm F}\rangle, 
\label{EE0}
\end{equation}
where we used $H\Psi_0=E_0\Psi_0$ and assumed that $\varepsilon$ is 
small and $\langle\psi_{\rm F}|H|\psi_{\rm F}\rangle \gg E_0$. 
The energy change is approximately estimated by the 
product of the probability, $\varepsilon^2$, of 
finding the forbidden component in $\Psi_0$ and 
the energy expectation value of the Pauli-forbidden state, 
$\langle\psi_{\rm F}|H|\psi_{\rm F}\rangle$. 
We show in Table~\ref{ENEDIFFQP} $\varepsilon^2$, 
$\langle\psi_{\rm F}|H|\psi_{\rm F}\rangle $, and the energy 
change calculated with the different potentials. 
In the case of $_\Lambda^5$He, for example, the set A potential 
with soft central repulsion gives $\varepsilon^2=0.00406$, 
$\langle\psi_{\rm F}|H|\psi_{\rm F}\rangle= 513$ MeV, 
while the set C potential with very large 
repulsion gives $\varepsilon^2=0.00331$, 
$\langle\psi_{\rm F}|H|\psi_{\rm F}\rangle=969$ MeV.
We see that these two 
factors really depend on the potential but their product is 
rather weakly dependent on the potential, and close to the 
binding energy change $\delta E$. This feature is a 
basic reason for the mild dependence of the energy change 
on the potential. 
A similar result is seen in the case of $_{\Lambda\Lambda}^{\ \ 5}$H, 
where the OBE-sim or FSS-sim $\Lambda\Lambda$ potential produces 
almost the same energy change though the core height is quite 
different between the two potentials. 
It is also interesting to recognize the difference of the energy change 
between $_\Lambda^5$He and $_{\Lambda\Lambda}^{\ \ 5}$H. 
Comparing the results calculated with set A and FSS-sim potentials  
we understand that, though $\langle\psi_{\rm F}|H|\psi_{\rm F}\rangle$ 
is almost the same in both cases, $\varepsilon^2$ is significantly 
different in both nuclei, which explains the larger quark Pauli effect in 
$_{\Lambda}^{5}$He than in $_{\Lambda\Lambda}^{\ \ 5}$H.  

\section{Summary}

We have determined phenomenologically the $\Lambda N$ central potential 
so as to reproduce the binding energies of $A=3,\, 4$ 
$\Lambda$-hypernuclei on the premise that various effects such as 
the noncentral forces and the $\Lambda$-$\Sigma$ conversion etc. 
are renormalized in the effective potential. 
The phase shifts predicted by this potential 
reaches 32 and 19 degrees at maximum in $^1S_0$ and $^3S_1$ states, and 
the scattering lengths and effective ranges are $a_s=-2.52$ fm, 
$a_t=-1.20$ fm, $r_s=3.08$ fm and $r_t=4.26$ fm, respectively. 

Bound state solutions for the Schr\"{o}dinger equation have been 
obtained by the stochastic variational 
method with the correlated Gaussian basis. We have performed a 
high precision calculation to determine the binding energies of 
the hypernuclei with $A \le 6$. We have properly taken into 
account both the short-range correlation of the particles and 
the asymptotic behavior of the wave function in setting up the 
variational trial functions. We have employed the three phase-equivalent 
$\Lambda N$ potentials which have different core heights and 
confirmed that they all give almost the same 
$\Lambda$ separation energies for the $s$-shell $\Lambda$-hypernuclei 
($_\Lambda^3$H, $_\Lambda^4$H, $_\Lambda^4$He, $_\Lambda^4$H$^\ast$, 
$_\Lambda^4$He$^\ast$, $_\Lambda^5$He). 

The charge symmetry breaking part of the $\Lambda N$ potential 
has been determined so as to reproduce both  
$\Delta B_\Lambda$ and $\Delta B_\Lambda^\ast$ values 
for the $A=4$ $\Lambda$-hypernuclei. 
We find out that the CSB is weakly spin-dependent 
in accordance with the conclusion of Ref.~\cite{Bod85}. 
The $\Lambda p$ potential is more attractive than the $\Lambda n$ 
one in both $^1S_0$ and $^3S_1$ states. 
The $\Lambda p$ phase shifts increase by 2 degrees with the inclusion of 
the CSB effect. 
This phenomenology disagrees 
with the prediction of some OBE models where the CSB is 
spin-dependent and 
the $\Lambda n$ potential is more attractive than the $\Lambda p$ 
potential in $^1S_0$ state. 

For the study of the $\Lambda\Lambda$-hypernuclei we have used the 
two different \LL potentials which reproduce 
the $\Delta B_{\Lambda\Lambda}(_{\Lambda\Lambda}^{\ \ 6}{\rm He})$ 
value reasonably well. The present calculation has shown that we can 
expect hitherto undiscovered particle-stable $\Lambda\Lambda$-hypernuclei, 
$_{\Lambda\Lambda}^{\ \ 4}$H, $_{\Lambda\Lambda}^{\ \ 5}$H and 
$_{\Lambda\Lambda}^{\ \ 5}$He, with the $B_{\Lambda\Lambda}$ 
values of 0.4, 5.5 and 6.3 MeV, respectively. No other bound  
$\Lambda\Lambda$-hypernucleus has been obtained for $A\le 5$. 
Since the binding energy of $_{\Lambda\Lambda}^{\ \ 4}$H is very small, 
its existence will crucially depend on the attraction of the 
$\Lambda\Lambda$ interaction, while both of $_{\Lambda\Lambda}^{\ \ 5}$H and 
$_{\Lambda\Lambda}^{\ \ 5}$He are more tightly bound, so that 
their existence will be very probable. 
Experimental confirmation of these $\Lambda\Lambda$-hypernuclei 
will provide us with 
valuable information on the $\Lambda\Lambda$ interaction. 

The structure change produced by adding $\Lambda$ particles has been 
investigated by calculating the density distribution, two-particle 
correlation function and spectroscopic factor etc. The distribution 
of $\Lambda$ is in general much broader than that of $N$. 
For instance, the root-mean-square distance of the two $\Lambda$ 
particles in $_{\Lambda\Lambda}^{\ \ 4}$H is larger than 8 fm and 
they spend almost all the time outside the range of the \LL 
interaction.

The present calculation has confirmed the well-known problem that 
the binding energies of both $_\Lambda^5$He and 
$_{\Lambda\Lambda}^{\ \ 6}$He are predicted to be overbound by 
$2-3$ MeV. We have focused our attention on the effect of the 
quark substructure of $N$ and $\Lambda$ to see if the binding energy 
is reduced by the quark Pauli principle. 
The effect is negligible if the baryon size in which the quarks are confined 
is smaller than 0.6 fm, 
but becomes appreciable, particularly in $_{\Lambda\Lambda}^{\ \ 6}$He, 
if the size is taken to be as large as 0.7 fm. To resolve this 
overbinding problem still remains very challenging and important 
in few-body hypernuclear systems.

\appendix
\section{Matrix elements of the correlated Gaussians}

Most of the formulas for the matrix elements of the CG 
are given elsewhere\cite{SVM98,CPC97,USU98,SU98}. 
The reader is referred to these references for the detail, 
so here we recapitulate the basic formulas and the 
calculation of the correlation function. 
We consider the system comprising $N$ particles and denote 
the set of relative coordinates as $({\bfi x}_1, \cdots, {\bfi x}_{N-1})$. 

The CG has two noteworthy advantages in 
the calculation of matrix elements 
in addition to its flexibility to represent various shapes of 
many-variable functions. First the CG does 
not change its form under the permutation of particle indices. 
Since the permutation $P$ induces a linear transformation of the 
coordinates as 

\begin{equation}
P{\bfi x}_i = \sum_{j=1}^{N-1} T_{ij}{\bfi x}_j\ \ \ \ \ (i=1,\cdots,N-1),
\end{equation}
we notice that the CG is changed by $P$ into 

\begin{equation}
PG({\bfi x}, A)=G({\bfi x}, \widetilde{T}AT)\ \ \ \ \ {\rm with}\ \ \  
\widetilde{T}_{ij}=T_{ji}.
\end{equation}
The action of $P$ is thus very simple: The CG remains a new CG with 
the matrix $A$ being replaced with a new one $\widetilde{T}AT$. Thus the 
antisymmetry requirement for the basis function involving the CG 
is relatively easily fulfilled 
by just knowing the matrix $T$. We do not need to calculate 
other matrix elements than those between the CGs. 
Secondly most of the matrix elements for the CG can be obtained analytically,  
and this makes it possible to obtain a precise solution with the CG basis. 

The overlap of the CG is given by 
\begin{eqnarray}
\langle G({\bfi x}, A)|G({\bfi x}, A^{\prime})\rangle 
= \left(\frac{(2\pi)^{N-1}}{\det{ B}} \right)^{3\over 2}\ \ \ \ \ 
{\rm with}\ \ \ B=A+ A^{\prime}. 
\label{CGOVL}
\end{eqnarray}
By expressing the kinetic energy operator as 
\begin{equation}
\sum_{i=1}^{N}{{\bfi p}_i^2\over 2m_i}-T_{cm}={1\over 2}
\sum_{i,j=1}^{N-1}\Lambda_{ij}{\bfi \pi}_i\cdot{\bfi \pi}_j\ \ \ \ \ 
{\rm with}\ \ \ {\bfi \pi}_j=-i\hbar{\partial \over \partial{\bfi x}_j}, 
\end{equation}
we obtain the matrix element for the kinetic energy operator: 
\begin{eqnarray}
& &\langle G({\bfi x}, A)|\sum_{i=1}^{N}{{\bfi p}_i^2\over 2m_i}-T_{cm}|
G({\bfi x}, A^{\prime})\rangle \nonumber \\
& & \quad =\, {3 \over 2}\hbar^2{\rm Tr}\left(A^{\prime}B^{-1}A\Lambda\right)
\langle G({\bfi x}, A)|G({\bfi x}, A^{\prime})\rangle.
\end{eqnarray}

To calculate the matrix element for the two-body interaction, it is 
convenient to express the potential in the form 
\begin{equation}
V({\bfi r}_i-{\bfi r}_j)=\int d{\bfi r} V({\bfi r})
\delta({\bfi r}_i-{\bfi r}_j-{\bfi r})
\end{equation}
and note that the relative distance vector of the particles can be 
written in terms of the coordinates {\bfi x} as 
\begin{equation}
{\bfi r}_i-{\bfi r}_j=\sum_{k=1}^{N-1}\omega_k {\bfi x}_k
=\tilde{\omega}{\bfi x}.
\end{equation}
Then the matrix element of the potential is given by 
\begin{eqnarray}
& & \langle G({\bfi x}, A)|V({\bfi r}_i-{\bfi r}_j)|
G({\bfi x}, A^{\prime})\rangle \nonumber \\
& & \quad =\, \int d{\bfi r} V({\bfi r})
\langle G({\bfi x}, A)|\delta(\tilde{\omega}{\bfi x}-{\bfi r})
|G({\bfi x}, A^{\prime})\rangle \nonumber \\
& & \quad =\, \left({c\over 2\pi}\right)^{3\over 2}
\langle G({\bfi x}, A)|G({\bfi x}, A^{\prime})\rangle
\int d{\bfi r} V({\bfi r}) 
\exp\left(-{1\over 2}cr^2\right),
\label{MEPOT}
\end{eqnarray}
where 
\begin{equation}
{1\over c}=\tilde{\omega}B^{-1}\omega.
\end{equation}
To prove this formula, one 
makes use of the Fourier integral for the $\delta$-function and the 
many-variable Gauss integration. The advantages of this method 
are that, since the dependence on the form of the potential appears only 
through the last integration, different types of potentials can be  
treated on equal footing and also that the correlation function 
(\ref{CRFN}) can already be calculated in the above procedure. 
By choosing, e.g., $V({\bfi r})=r^n$ we can get the matrix element of 
$|{\bfi r}_i-{\bfi r}_j|^n$ very easily. 

The calculation of the density distribution of the particle as defined by 
Eq.~(\ref{RHOFN}) poses no problem if 
one notices that ${\bfi r}_i-{\bfi R}_c$ is expressed in terms of 
a linear combination of $({\bfi x}_1,\cdots,{\bfi x}_{N-1})$. 
One only needs to redefine $\omega$ appropriately and 
then use the formula of Eq.~(\ref{MEPOT}).

By generalizing the method presented above for the matrix element of the 
$\delta$-function, we can calculate the two-particle distribution 
function (\ref{TPDFN}) or even more-particle distribution function. 
All of these functions can be obtained through the following 
formula:
\begin{eqnarray}
&&\langle G({\bfi x},A)|
\prod_{i=1}^{n}\delta(\widetilde{{w}^{(i)}}{\bfi x}-{\bfi \rho}_i)
|G({\bfi x},A^{\prime})\rangle \nonumber \\
& & \quad = \, 
\left({{\rm det}{\cal C}\over (2\pi)^n}\right)^{3\over 2}
\exp\left(-{1\over 2} \sum_{i,j=1}^n {\cal C}_{ij}
{\bfi \rho}_i\cdot{\bfi \rho}_j\right) \nonumber \\
& & \quad \ \times \, \langle G({\bfi x}, A)|G({\bfi x}, A^{\prime})\rangle, 
\end{eqnarray}
where ${\cal C}$ is an $n\times n$ matrix defined by 
\begin{equation}
({\cal C}^{-1})_{ij}=\widetilde{{w}^{(i)}}{B}^{-1}w^{(j)}. 
\end{equation}
By integrating the above equation 
over ${\bfi \rho}_1,\cdots,{\bfi \rho}_n$, one recovers 
the overlap (\ref{CGOVL}) of the CG.

\clearpage

\begin{figure}
\caption{The $^1S_0$ (left) and $^3S_1$(right) phase shifts of 
   $\Lambda N$ scattering as a function of $\Lambda$ momentum 
   $p_{\Lambda}$. 
   Solid lines are obtained by the potentials of sets A, B and C, 
   $+$ by FSS, and $\times$ are by the Nijmegen model F, respectively. 
   \label{LNPHSFT}} 

\caption{The $\Lambda N$ potentials of sets A and C as a function of 
   $N$-$\Lambda$ distance $r$. 
    \label{LNPOTSHP}}

\caption{Forward-backward ratio of $\Lambda p$ scattering as a function of 
   $\Lambda$ momentum $p_{\Lambda}$.
          Experimental data are taken from Refs.~\protect\cite{LpAlex} 
          and \protect\cite{LpSech}. 
    \label{LNFB}} 

\caption{Total cross section of $\Lambda p$ scattering as a function of 
 $\Lambda$ momentum $p_{\Lambda}$. The dotted line is 
  obtained with the potentials of sets A, B and C, while the solid line 
  is obtained with set A potential including CSB. 
  Experimental data are taken from 
 Refs.~\protect\cite{LpAlex,LpSech,LpKady,LpHaup}.
\label{LPTOT} }

\caption{The $^1S_0$ phase shifts (left) calculated by the \LL potentials 
 displayed on the right panel. 
\label{LLPHSFTnPOTSHP}}

\caption{The energy convergence of $^{\ \ 5}_{\Lambda\Lambda}$He 
as a function of the basis dimension $K$. 
Set A \LN and FSS-sim \LL potentials are used. The CSB effect as we discuss 
in subsect.\,3.4 is included. 
The upper panel is the step-by-step selection up to $K$=200, and 
the lower panel shows the refinement at $K$=200, which is followed by 
the step-by-step basis extension to $K$=300 and 
the second refinement at $K$=300. 
\label{h5llconv}}

\caption{The density distributions of $N$ and $\Lambda$ 
for $^{2}$H, $_{\Lambda}^{3}$H and $_{\Lambda\Lambda}^{\ \ 4}$H 
as a function of $r$, the distance from the center-of-mass of $^{2}$H. 
\label{h4ll}}

\caption{The density distributions of $N$ and $\Lambda$ 
for $^{3}$H, $_{\Lambda}^{4}$H, $_{\Lambda}^{4}$H$^*$  
and $_{\Lambda\Lambda}^{\ \ 5}$H 
as a function of $r$, the distance from the center-of-mass 
of $^{3}$H.
\label{h5ll}}

\caption{The density distributions of $N$ and $\Lambda$ 
for $^{4}$He, $_{\Lambda}^{5}$He and $_{\Lambda\Lambda}^{\ \ 6}$He 
as a function of $r$, the distance from the center-of-mass of 
$^{4}$He. 
\label{h6ll}}

\caption{The correlation function of two $\Lambda$ particles 
for the $\Lambda\Lambda$-hypernuclei, as a function of $r$, 
the distance between $\Lambda$-$\Lambda$. 
\label{cfll}}

\caption{The two-$\Lambda$ distribution function 
$r^2r^{\prime 2}\sin\theta D({\bfi r},{\bfi r}^\prime)$ (upper) of 
$^{\ \ 5}_{\Lambda\Lambda}$H as a function of $r$ and $\theta$, 
where $r^\prime=1.8$ fm and $(x,y)=(r\cos\theta,r\sin\theta)$. 
The core nucleus $^3$H is located at the origin and 
one of the two $\Lambda$ particles is placed at 
the point denoted X. 
The lower panel plots the contour line of the distribution function. 
Set A \LN and OBE-sim \LL potentials are used. 
\label{tpden}}
\end{figure}


\clearpage

\begin{figure}
 \begin{minipage}[t]{60mm} 
	\epsfxsize=7cm
	\epsfbox{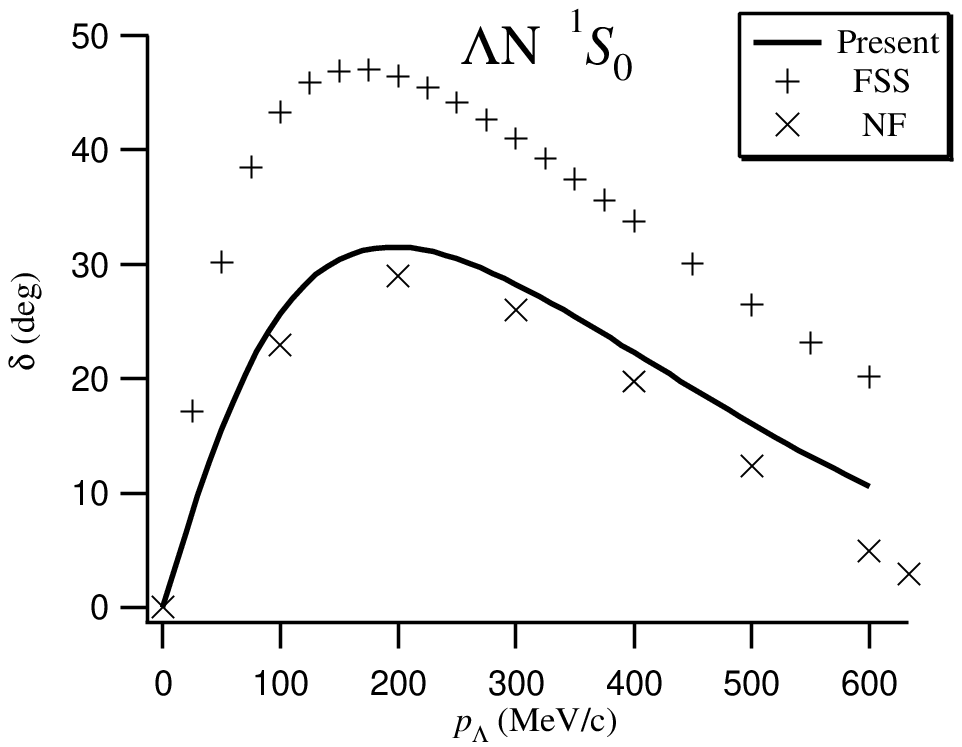}
 \end{minipage}
\hspace{\fill}
 \begin{minipage}[t]{60mm}
	\epsfxsize=7cm
	\epsfbox{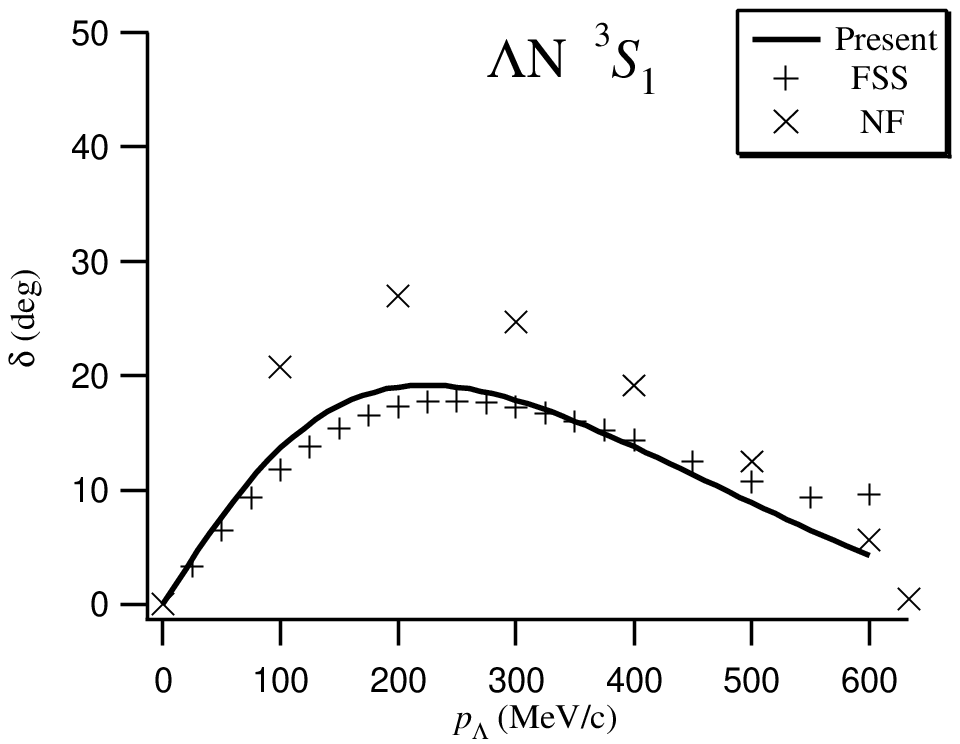}
 \end{minipage}\\
Fig. 1
\end{figure}
\bigskip

\begin{figure}
    \centering \leavevmode 
	\epsfbox{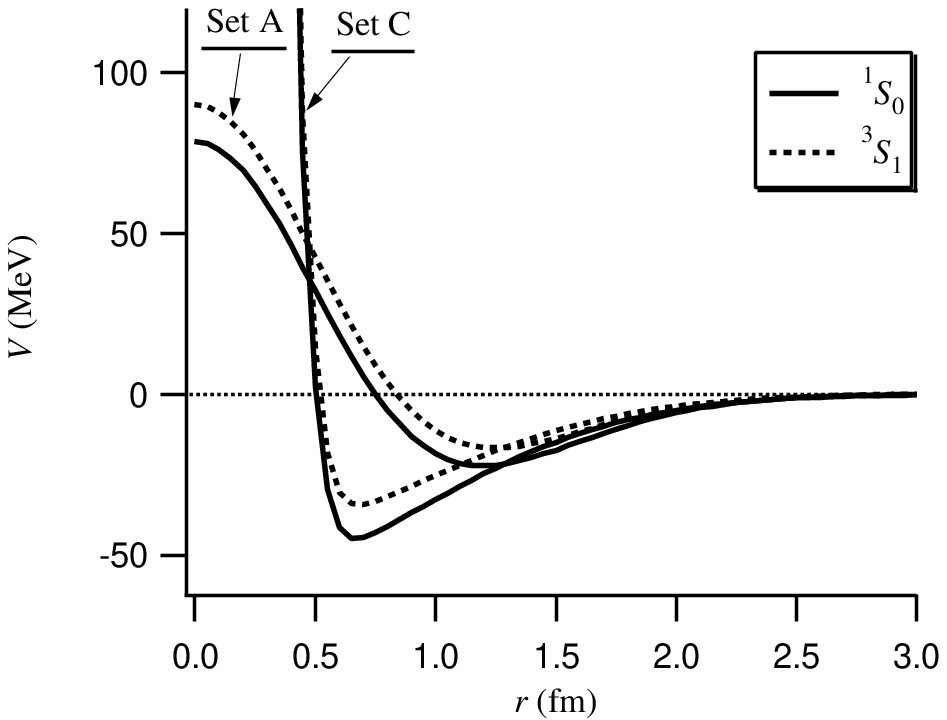}\\
Fig. 2
\end{figure}
\bigskip

\begin{figure}
    \centering \leavevmode 
	\epsfbox{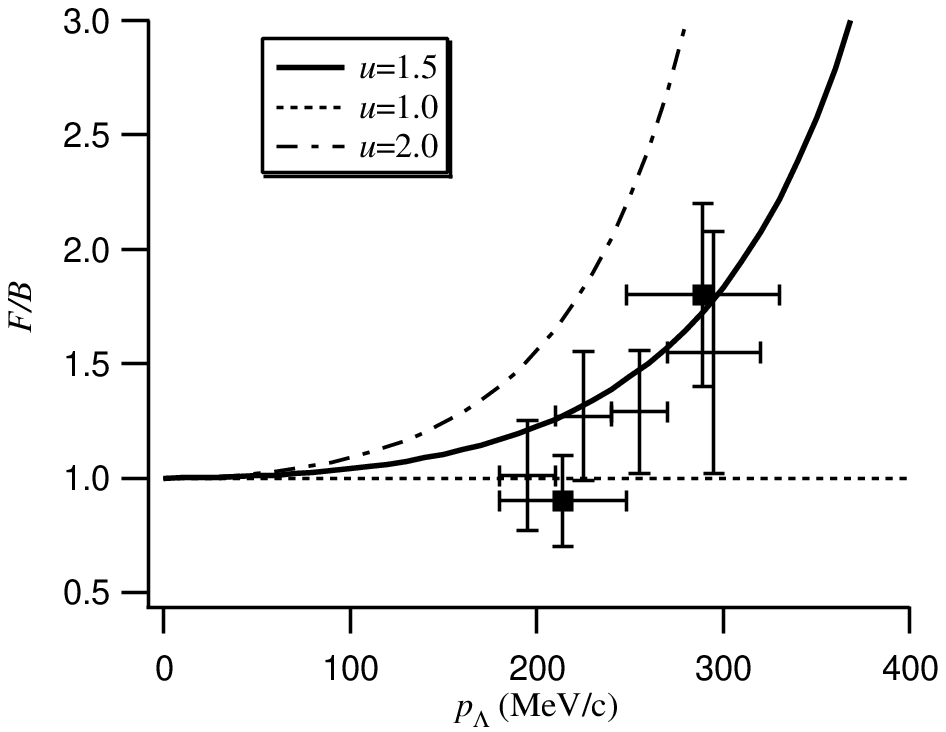}\\
Fig. 3
\end{figure}
\bigskip

\begin{figure}
    \centering \leavevmode
	\epsfbox{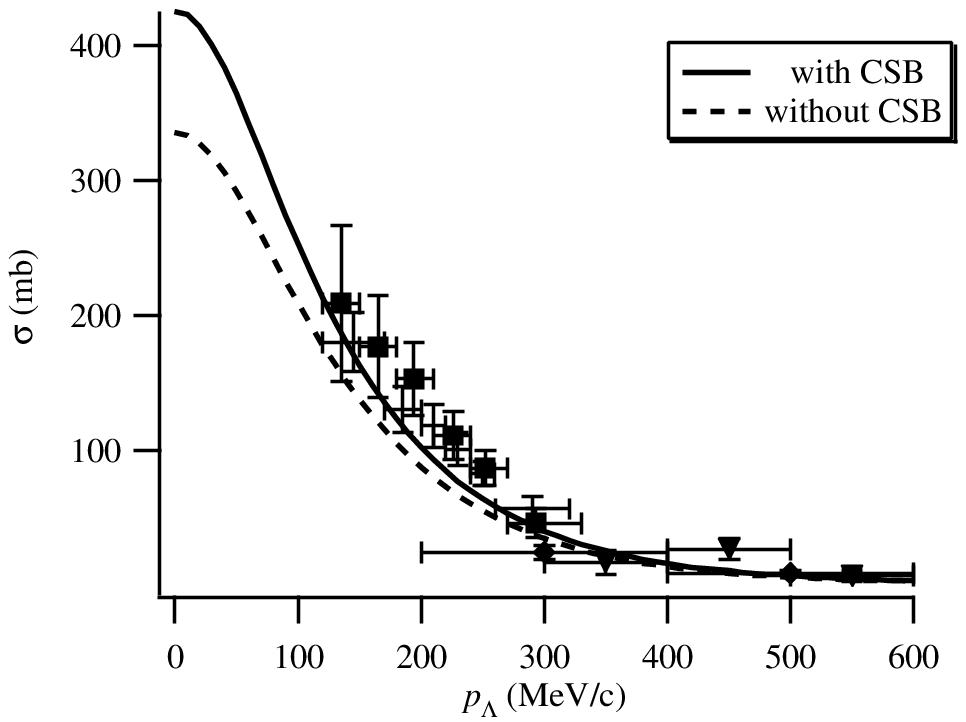}\\
Fig. 4
\end{figure}
\bigskip

\begin{figure}
 \begin{minipage}[t]{60mm} 
	\epsfxsize=7cm
	\epsfbox{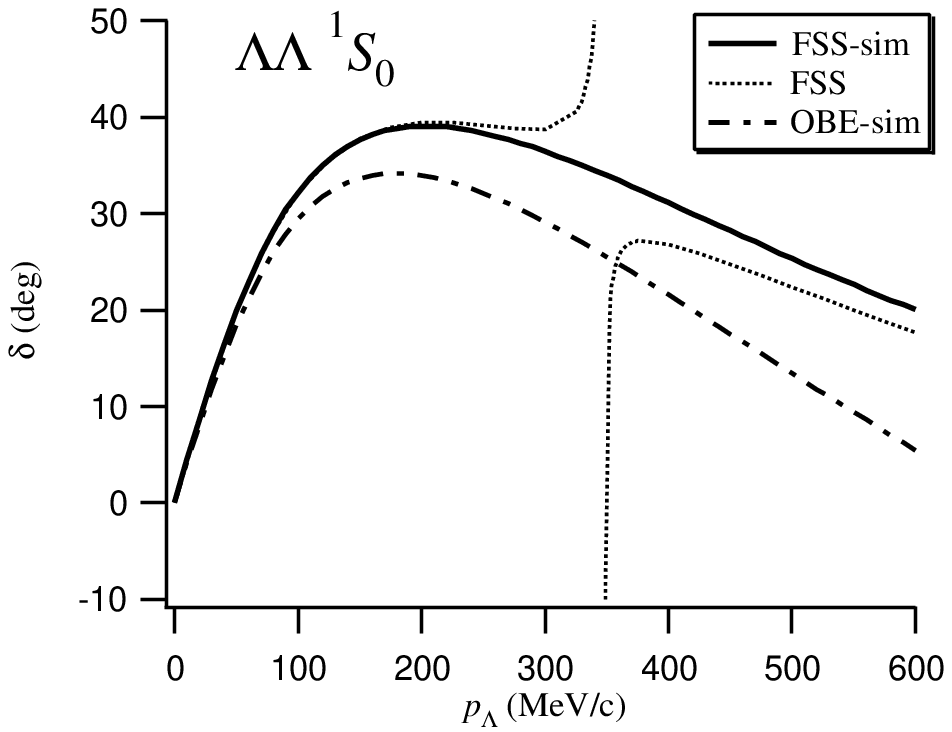}
 \end{minipage}
\hspace{\fill}
 \begin{minipage}[t]{60mm}
	\epsfxsize=7cm
	\epsfbox{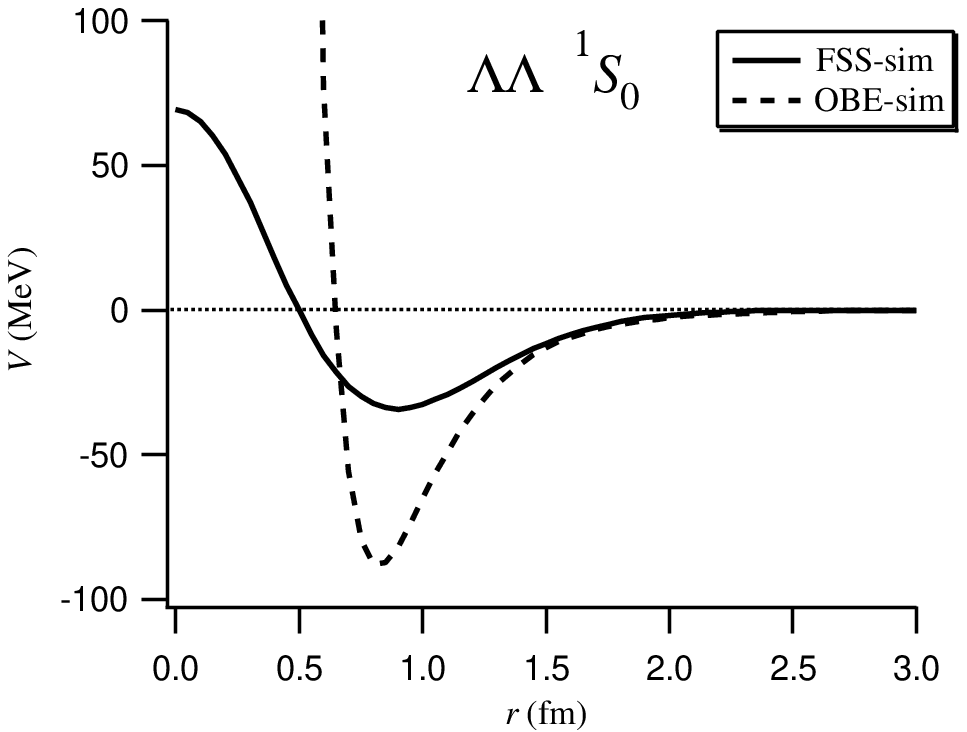}
 \end{minipage}\\
Fig. 5
\end{figure}
\bigskip

\begin{figure}
 \noindent 
 \centering \leavevmode  \epsfxsize=10cm 
 \epsfbox{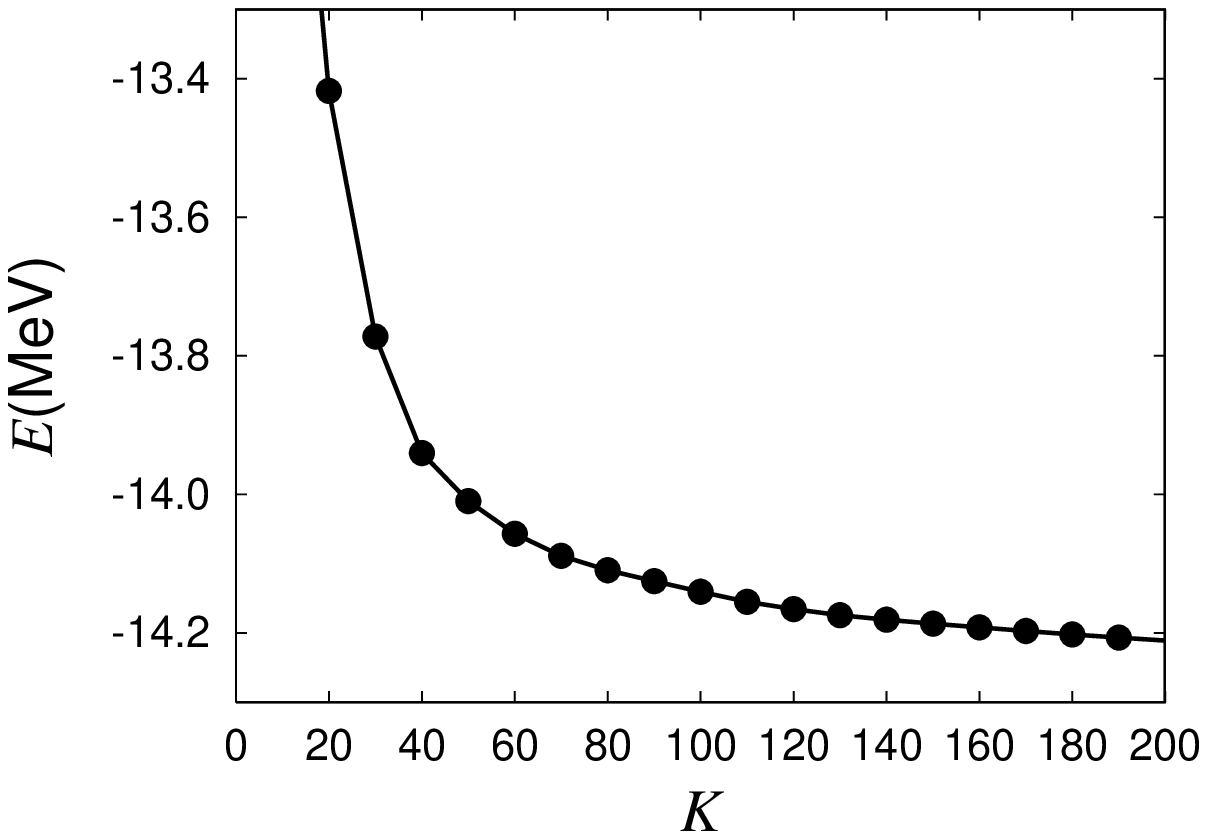} \\ 
 \centering \leavevmode  \epsfxsize=10cm 
 \epsfbox{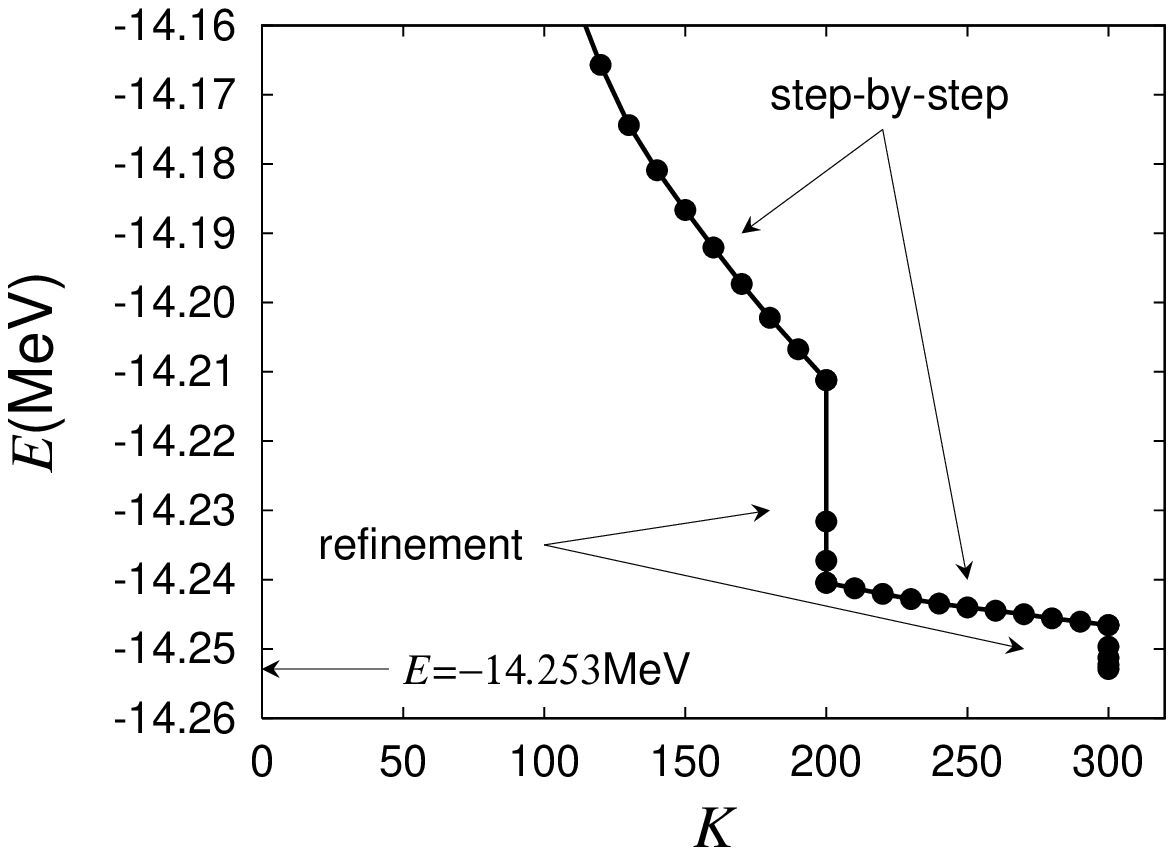} \\
Fig. 6
\end{figure}
\bigskip

\begin{figure}[htbp]
\centering \leavevmode
	\epsfbox{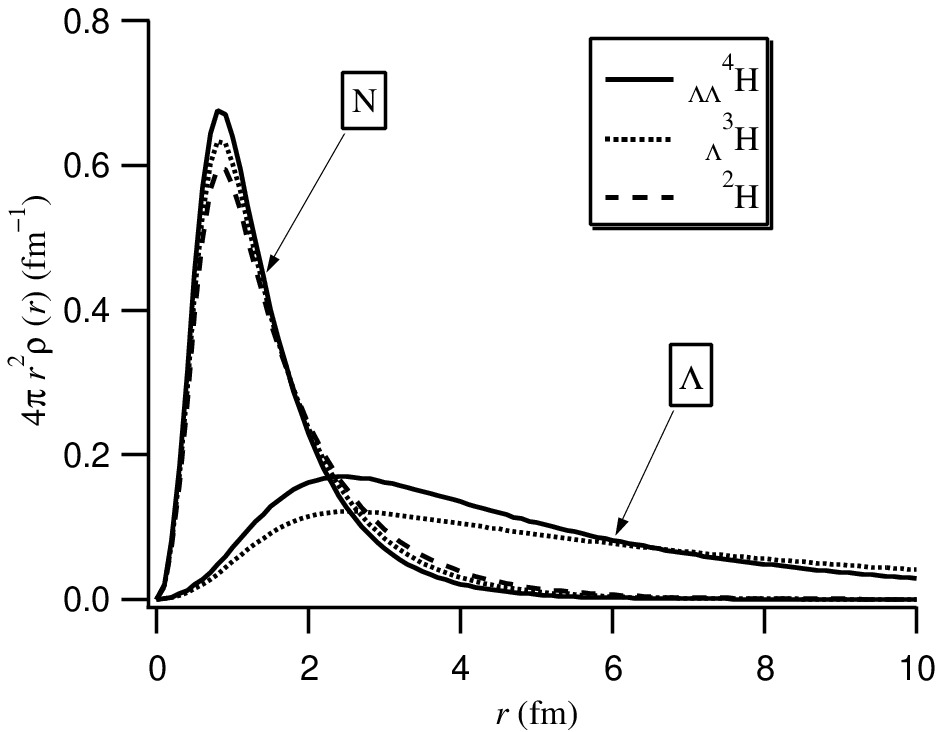}\\
Fig. 7
\end{figure}
\bigskip

\begin{figure}[htbp]
\centering \leavevmode
	\epsfbox{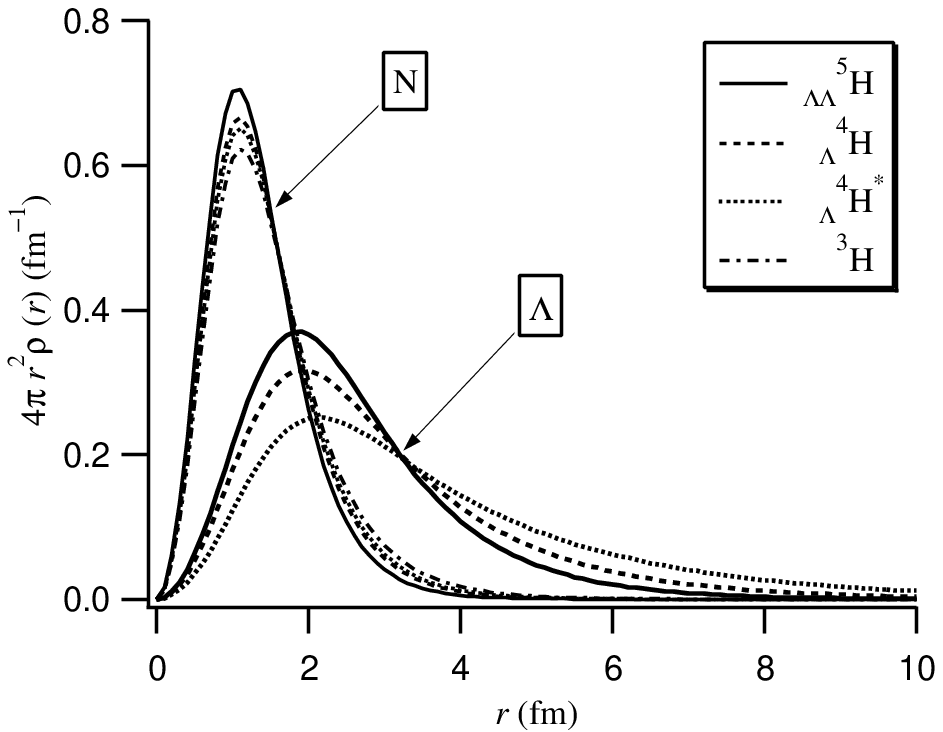}\\
Fig. 8
\end{figure}
\bigskip

\begin{figure}[htbp]
\centering \leavevmode
	\epsfbox{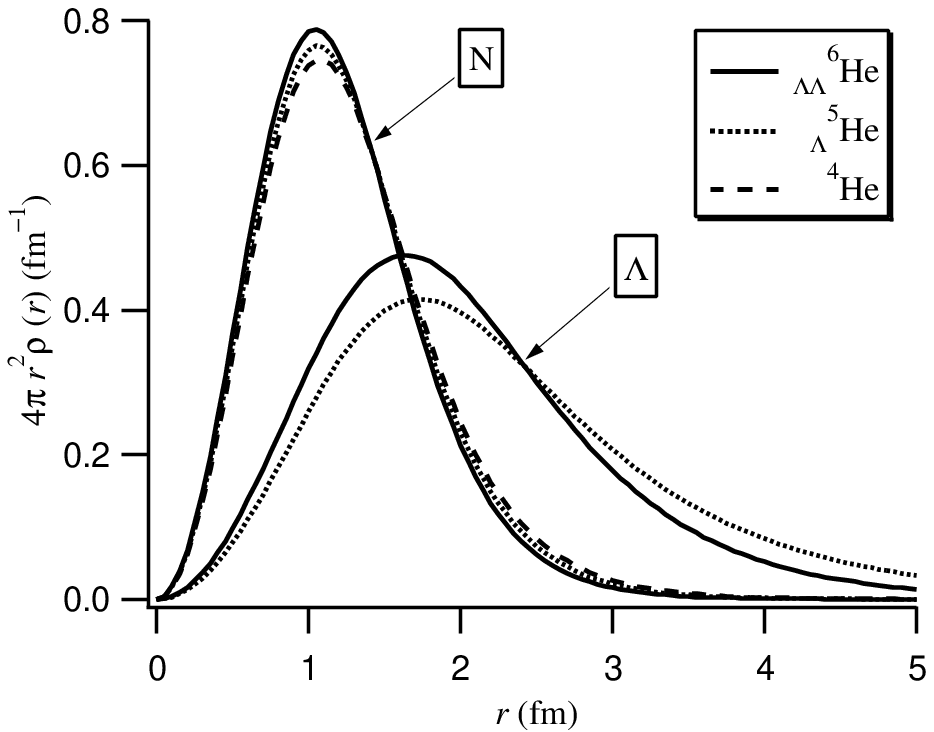}\\
Fig. 9
\end{figure}
\bigskip

\begin{figure}[htbp]
\centering \leavevmode
	\epsfbox{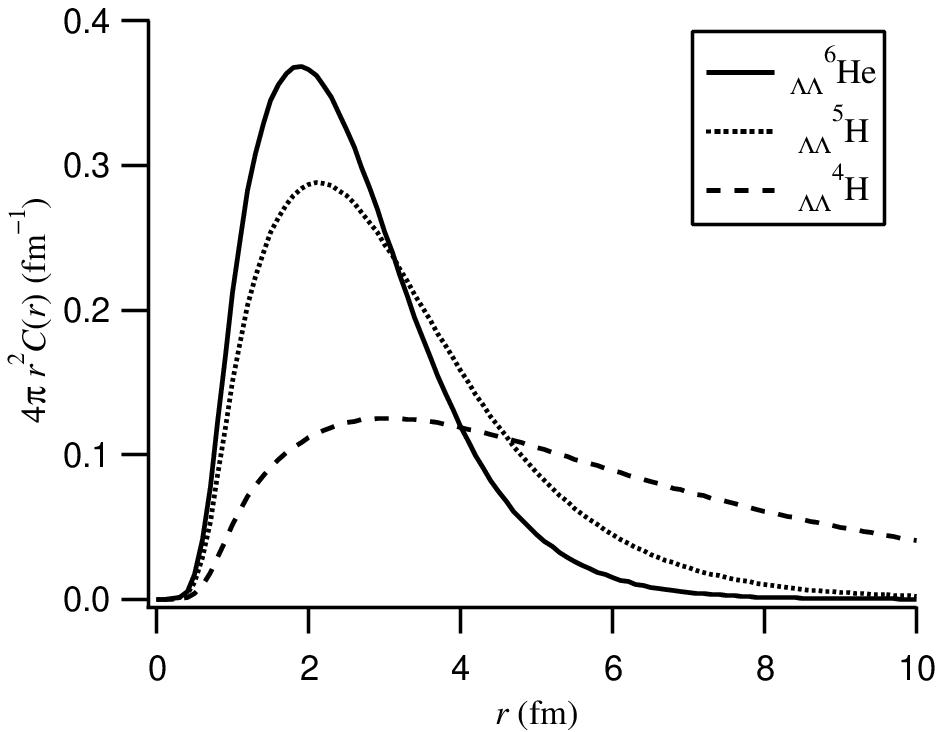}\\
Fig. 10
\end{figure}
\bigskip

\begin{figure}[htbp]
	\centering \leavevmode
	\epsfxsize=12cm 
	\epsfbox{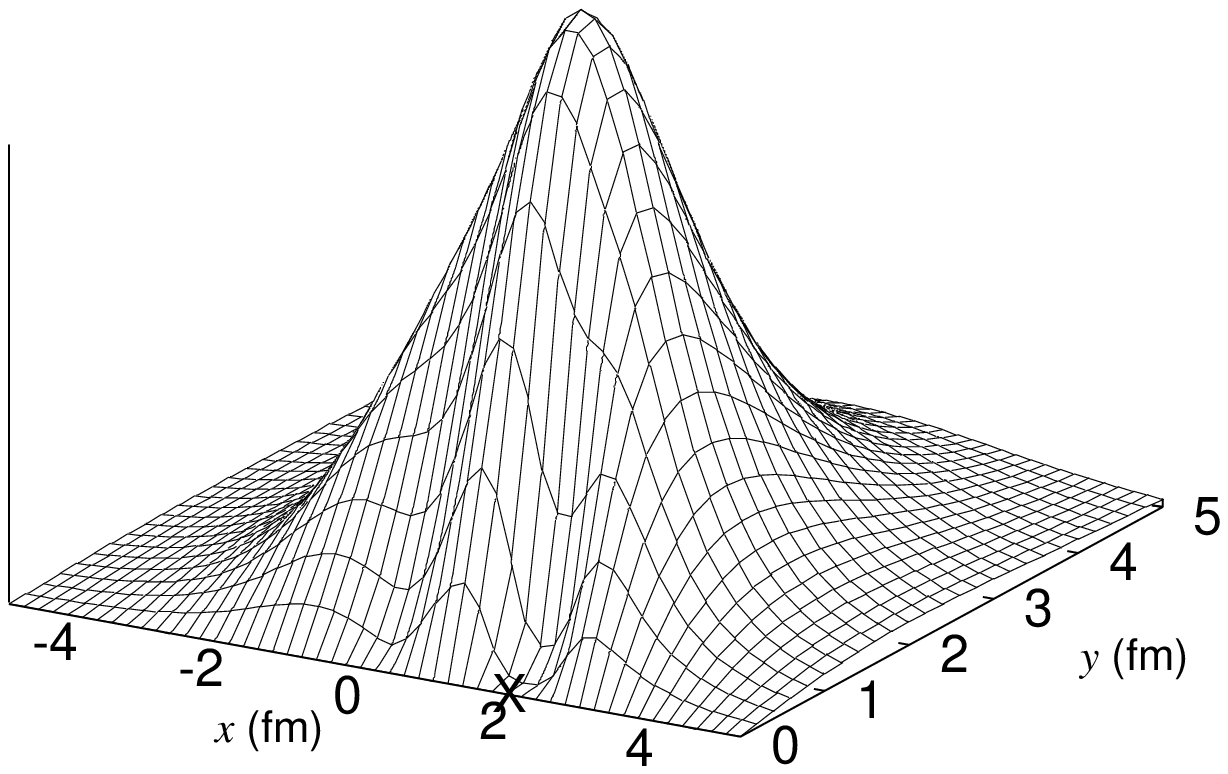}\\ 
\hspace{-4.0cm} 
	\epsfbox{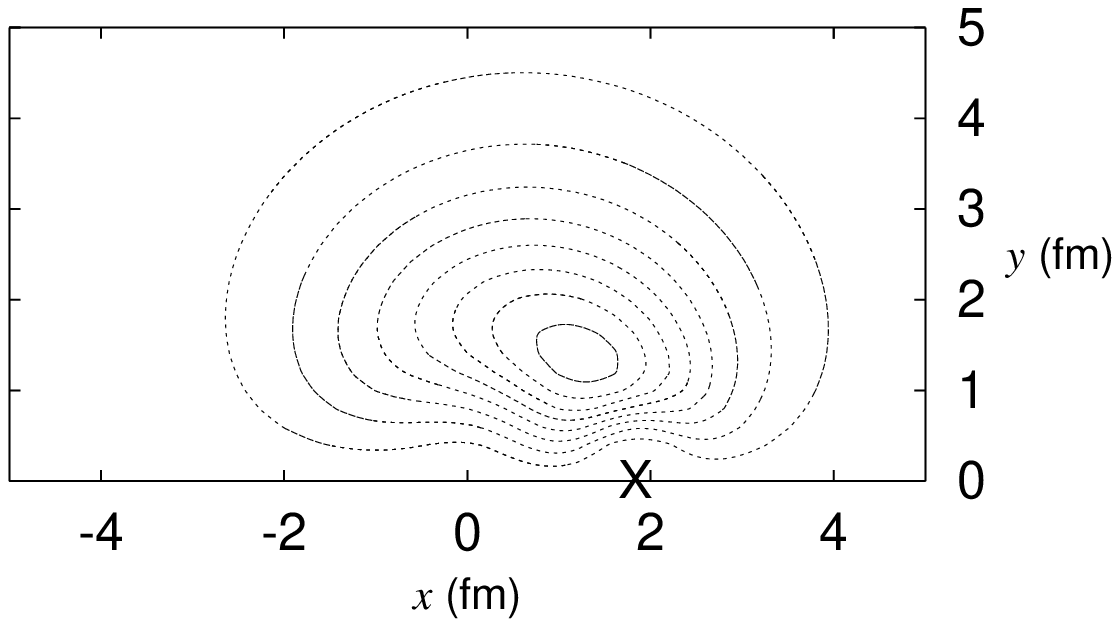}\\
Fig. 11
\end{figure}
\bigskip

\clearpage

\begin{table} \centering \leavevmode
 \caption{Contributions of $^1S_0$ and $^3S_1$ \LN 
     potentials to the $\Lambda$ separation energy in a 
     core-nucleus+$\Lambda$ model.  
     $J$ is the total angular momentum and $J_c$ is the angular momentum 
     of the core nucleus. 
     \label{TSWGTS}}
  \begin{tabular}{ccccc}
    \hline 
 	    & $J$ & $J_c$ & $N_s$ & $N_t$ \\
    \hline
	\Lower{$^3_{\Lambda}$H} & $1\over 2$ & 1 & $3\over 2$ & $1\over 2$ \rule{0in}{3ex}\\[.5ex]
    \cline{2-5}
	  & $3\over 2$ & 1 &	0	& 2 \rule{0in}{3ex}\\[.5ex]
    \hline
	\Lower{ $^4_{\Lambda}$H, $^4_{\Lambda}$He} & 0 & $1\over 2$ & 
    $3\over 2$ & $3\over 2$ \rule{0in}{3ex}\\[.5ex]
    \cline{2-5}
	  & 1 & $1\over 2$ & $1\over 2$ & $5\over 2$ \rule{0in}{3ex}\\[.5ex]
    \hline
	$^5_{\Lambda}$He & $1\over 2$ & 0 & 1	& 3 \rule{0in}{3ex}\\[.5ex]
    \hline 
  \end{tabular}
\end{table}

\begin{table} \centering \leavevmode 
 \caption{Parameters of $NN$, \LN and \LL potentials. 
     Three sets for the \LN potential are charge symmetric, giving the 
    same $^1S_0$ and $^3S_1$ phase shifts in low energy region 
    ($p_\Lambda\le600$ MeV/c): 
     The scattering length and effective range are $a_s = -2.52$ fm, 
    $r_s = 3.08$ fm, and $a_t = -1.20$ fm,  $r_t = 4.26$ fm. 
    To include the charge symmetry breaking for set A \LN potential, increase  
    (decrease) both $V_{0t}$ and $V_{0s}$ by the values denoted CSB for 
    $\Lambda p$ ($\Lambda n$). 
    \label{LNPOTPARA} }
  \begin{tabular}{llcccccc}
    \hline 
 	& & $V_{0R}$  & $V_{0t}$  & $V_{0s}$  & $\kappa_R$  & $\kappa_t$  & 
	$\kappa_s$  \\
 	& & (MeV) & (MeV) & (MeV) & (fm$^{-2}$) & (fm$^{-2}$) & (fm$^{-2}$) \\
    \hline
	$NN$ & Ref.\cite{Min77} & 
			200.0 & 178.0 & 91.85 & 1.487 & 0.639 & 0.465 \\
    \hline
	& Set A & 200.0 & 109.8 & 121.3 & 1.638 & 0.7864 & 0.7513 \\
	\LN & CSB$\{^{\Lambda p}_{\Lambda n}\}$ & 
	& $\{\pm 3.3\}$ & $\{\pm 2.7\}$ & & & \\
    \cline{2-8}
	& Set B &	600.0 &	52.61 &	66.22 &	5.824 &	0.6582 & 0.6460 \\
    \cline{2-8}
	& Set C &	5000.0 &47.87 &	61.66 &	18.04 &	0.6399 & 0.6325 \\
    \hline
	\LL & FSS-sim &	200.0 &--& 130.8 & 2.776 &-- & 1.062 \\
    \hline 
  \end{tabular}
\end{table}

\begin{table} \centering \leavevmode 
\caption{Values of $d_{max}$, given in units of fm, and the number 
of spin functions $g$ used in the present calculation. 
All possible spin functions are taken into account except for 
$^{\ \ 6}_{\Lambda\Lambda}$He, which is restricted to  
$[[[S_p,S_p]_0,[S_n,S_n]_0]_0,[S_\Lambda,S_\Lambda]_0]_{00}$. 
\label{PARAM}}
 \begin{tabular}{lcc}
  \hline 
 &	$d_{max}$ &	$g$ \\
  \hline
$\ \ ^2$H & 20 &	1 \\
$\ \ ^3$H & 15 &	2 \\
$\ \ ^3$He & 15 &	2 \\
$\ \ ^4$He & 12 &	2 \\
$\ \ ^3_{\Lambda}$H & 45 &	2 \\
$\ \ ^4_{\Lambda}$H, $^4_{\Lambda}$He & 18 &	2 \\
$\ \ ^4_{\Lambda}$H$^{\ast}$, $^4_{\Lambda}$He$^{\ast}$ & 24 &	3 \\
$\ \ ^5_{\Lambda}$He &	15 &	5 \\
$^{\ \ 4}_{\Lambda\Lambda}$He &	55 &	3 \\
$^{\ \ 5}_{\Lambda\Lambda}$H, $^{\ \ 5}_{\Lambda\Lambda}$He & 18 &	5 \\
$^{\ \ 6}_{\Lambda\Lambda}$He & 15 &	1 \\
  \hline 
 \end{tabular}
\end{table}

\begin{table} \centering \leavevmode 
 \caption{The root-mean-square distances of $\Lambda$- and 
$\Lambda\Lambda$-hypernuclei. $K$ is the basis dimension and 
$\eta$ is the virial ratio defined by Eq.~(\protect\ref{VIRIAL}). 
   Set A \LN and OBE-sim \LL potentials are used. 
\label{rmstab}}
  \begin{tabular}{ccccccc}
   \hline 
 & $\sqrt{\langle r^2\rangle}$ & $\sqrt{\langle r_{NN}^2\rangle}$ & 
	$\sqrt{\langle r_{\Lambda N}^2\rangle}$ & 
	$\sqrt{\langle r_{\Lambda\Lambda}^2\rangle}$ & $K$ & $\eta$ \\
   \hline
$^2$H &		1.95 &	3.90 &	&	&	30 &	$2\times 10^{-13}$ \\
$^3$H &		1.71 &	2.95 &	&	&	30 &	$1\times 10^{-5}$ \\
$^3$He &	1.74 &	3.01 &	&	&	30 &	$1\times 10^{-4}$ \\
 &		&	&	&	&	60 &	$6\times 10^{-6}$ \\
$^4$He &	1.41 &	2.30 &	&	&	60 &	$1\times 10^{-5}$ \\
   \hline
$^3_{\Lambda}$H &4.9 &	3.6 &	10. &	&	30 &	$3\times 10^{-5}$ \\
$^4_{\Lambda}$H &2.0 &	2.7 &	3.8 &	&	100 &	$6\times 10^{-5}$ \\
$^4_{\Lambda}$H$^{\ast}$ &2.4 &	2.8 &	4.7 &	&100 &	$2\times 10^{-4}$ \\
 &		&	&	&	&	200 &	$2\times 10^{-5}$ \\
$^4_{\Lambda}$He &	2.0 &	2.8 &	3.8 &	&100 &	$1\times 10^{-4}$ \\
$^4_{\Lambda}$He$^{\ast}$ &2.4 &2.9 &	4.8 &	&200 &	$2\times 10^{-5}$ \\
$^5_{\Lambda}$He &	1.6 &	2.3 &	3.0 &	&200 &	$6\times 10^{-5}$ \\
   \hline
$^{\ \ 4}_{\Lambda\Lambda}$H &	4.2 &3.3 &7.1 &8.4 &200 &$5\times 10^{-4}$ \\
$^{\ \ 5}_{\Lambda\Lambda}$H &2.0 &2.6 &3.2 &	3.5 &200 &$2\times 10^{-4}$ \\
$^{\ \ 5}_{\Lambda\Lambda}$He &	2.0 &2.6 &3.3 &3.5 &200 &$2\times 10^{-4}$ \\
$^{\ \ 6}_{\Lambda\Lambda}$He &	1.6 &2.2 &2.6 &	2.8 &100 &$2\times 10^{-4}$ \\
 &		 &	 &	 &		&200 &	$8\times 10^{-5}$ \\
\hline 
\end{tabular}
\end{table}

\begin{table} \centering \leavevmode 
 \caption{$\Lambda$ separation energies and 
    excitation energies, given in units of MeV, 
    of $A=3-5$ $\Lambda$-hypernuclei. $a)$ Ref.~\protect\cite{BL73}, 
    $b)$ Ref.~\protect\cite{Ex79}.  
  \label{BEOFSL}}
  \begin{tabular}{lcccccc}
   \hline 
	  & 
	$B_\Lambda(^3_\Lambda\mbox{H})$ & $B_\Lambda(^4_\Lambda\mbox{H})$ & 
	$B_\Lambda(^4_\Lambda\mbox{He})$ & $B_\Lambda(^5_\Lambda\mbox{He})$ & 
	$E_{\mbox{x}}(_\Lambda^4\mbox{H}^*)$ & 
	$E_{\mbox{x}}(_\Lambda^4\mbox{He}^*)$ \\
   \hline
	Set A &	0.18 & 2.24 & 2.20 & 4.98 & 1.14 & 1.13 \\
   	with CSB &   0.18 & 2.05 & 2.40 & 4.98 & 1.08 & 1.19 \\
   \hline
	Set C &	0.17 & 2.21 & 2.18 & 4.90 & 1.13 & 1.12 \\
   \hline
	Expt.  &	$0.13\pm0.05^a$ & $2.04\pm0.04^a$ & $2.39\pm0.03^a$ & 
			$3.12\pm0.02^a$ & $1.04\pm0.04^b$ & $1.15\pm0.04^b$ \\
   \hline 
  \end{tabular}
\end{table}

\begin{table} \centering \leavevmode 
 \caption{$\Lambda\Lambda$ separation energies, given in units of MeV, 
    of $\Lambda\Lambda$-hypernuclei. $a)$ Ref.~\protect\cite{BLL66}. 
   \label{BEOFDL}}
  \begin{tabular}{ccccc}
   \hline 
	 & 
	$B_{\Lambda\Lambda}(_{\Lambda\Lambda}^{\ \ 4}\mbox{H})$ & 
	$B_{\Lambda\Lambda}(_{\Lambda\Lambda}^{\ \ 5}\mbox{H})$ & 
	$B_{\Lambda\Lambda}(_{\Lambda\Lambda}^{\ \ 5}\mbox{He})$ & 
	$B_{\Lambda\Lambda}(_{\Lambda\Lambda}^{\ \ 6}\mbox{He})$ \\
   \hline
	Set A,  OBE-sim &	0.41 & 5.6 & 5.5 & 14.3 \\
	with CSB 	&	0.41 & 5.2 & 6.0 & 14.3 \\
   \hline
	Set A, FSS-sim &	0.53 & 6.1 & 6.0 & 15.1 \\
	with CSB 	&	0.53 & 5.7 & 6.5 & 15.1 \\
   \hline
        Set C, OBE-sim &	0.39 & 5.5 & 5.4 & 13.9 \\
   \hline
	Expt. & & & & $10.9\pm0.6^a$ \\
   \hline 
  \end{tabular}
\end{table}

\begin{table} \centering \leavevmode 
 \caption{The spectroscopic factors. Set A \LN and OBE-sim \LL potentials 
are used.  
\label{SPFACTOR}}
  \begin{tabular}{ccccccc}
   \hline 
$_\Lambda^3$H & $_\Lambda^4$H (0$^+$) & $_\Lambda^4$H (1$^+$) & $_\Lambda^5$He 
& $_{\Lambda\Lambda}^{\ \ 4}$H & $_{\Lambda\Lambda}^{\ \ 5}$H & $_{\Lambda\Lambda}^ {\ \ 6}$He \\
   \hline
0.991 & 0.986 & 0.992 & 0.994 & 0.944 &	0.962 &	0.985 \\
\hline 
  \end{tabular}
\end{table}

\begin{table} \centering \leavevmode 
 \caption{Scattering lengths, in units of fm, predicted 
   by various $\Lambda N$ interactions including CSB.  
  \label{SCTLNGOFCSB}}
  \begin{tabular}{lccccc}
   \hline 
   & \multicolumn{2}{c}{$a_s$} & & \multicolumn{2}{c}{$a_t$}\\
   \cline{2-3} \cline{5-6}
   & $\Lambda p$ & $\Lambda n$ & & 	$\Lambda p$ & $\Lambda n$ \\
   \hline
   Present &	$-$2.83 &	$-$2.26 &	& $-$1.36 &	$-$1.06 \\
   NSC97f\cite{NSC97} &	$-$2.51 &	$-$2.68 && $-$1.75 &$-$1.67 \\
   NSC89\cite{NSC89} &	$-$2.73 & $-$2.86 & & $-$1.48 & $-$1.24 \\
   NF\cite{NF79} &		$-$2.18 & $-$2.40 & & $-$1.93 & $-$1.84 \\
   ND\cite{ND77} &$-1.77\pm 0.28$ & $-2.03\pm 0.32$ & & $-2.06\pm 0.12$ & $-1.84\pm 0.10$ \\
   Ref.~\cite{Nij73} &$-2.16\pm 0.26$ &	$-2.67\pm 0.35$ & &$-1.32\pm 0.07$ &	$-1.02\pm 0.05$ \\
   \Lower{Ref.~\cite{DHT72}
	$^{({\rm with}\ {\bfi \sigma}_\Lambda\cdot{\bfi \sigma}_N)}
	_{({\rm without}\ {\bfi \sigma}_\Lambda\cdot{\bfi \sigma}_N)}$} & 
	$-$1.83 & $-$2.45 & & $-$1.77 & $-$1.61 \\
 &	$-$2.45 & $-$1.83 & & $-$1.94 & $-$1.47 \\
   \hline 
\end{tabular}
\end{table}

\begin{table} \centering \leavevmode 
\caption{$\Lambda$ and $\Lambda\Lambda$ separation energies, 
given in units of MeV, of $A=5$ and $6$ hypernuclei 
as a function of the baryon size $b$ (in fm). 
The case of $b=0$ is no quark Pauli calculation. 
Set A $\Lambda N$ and OBE-sim $\Lambda\Lambda$ potentials are used. 
$a)$ Ref.~\protect\cite{BL73}, $b)$ Ref.~\protect\cite{BLL66}.
\label{SEPENEQP}}
\begin{tabular}{lccccc}
\hline 
   & \multicolumn{4}{c} {Theory} & Expt. \\
\cline{2-5}
   & {$b=0$} & 0.6 & 0.7 & 0.86 &  \\
\hline
$B_{\Lambda}(^5_{\Lambda}$He) & 
4.98 & 4.90 & 4.70 & 3.64 & 3.12$\pm$0.02$^a$  \\
\hline
$B_{\Lambda \Lambda}(^{\ \, 5}_{\Lambda \Lambda}$H) & 
5.6 & 5.6 & 5.5 & 5.1 & --\\
$B_{\Lambda \Lambda}(^{\ \, 5}_{\Lambda \Lambda}$He) & 
5.5 & 5.5 & 5.5 & 5.1 & --\\ 
\hline
$B_{\Lambda \Lambda}(^{\ \, 6}_{\Lambda \Lambda}$He) & 
14.3 & 13.9 & 13.0 & 9.5 & $10.9\pm 0.6^b$ \\ 
\hline 
\end{tabular}
\end{table}

\begin{table} \centering \leavevmode 
 \caption{$\Lambda$ and $\Lambda\Lambda$ separation energies (in MeV) 
   for different potentials. The baryon size is set to $b=0.86$ fm. 
   \label{SEPENEQP2}}
  \begin{tabular}{ccccc}
   \hline 
	 & 
	$B_{\Lambda}(_{\Lambda}^{5}\mbox{He})$ & 
	$B_{\Lambda\Lambda}(_{\Lambda\Lambda}^{\ \ 5}\mbox{H})$ & 
	$B_{\Lambda\Lambda}(_{\Lambda\Lambda}^{\ \ 6}\mbox{He})$ \\
   \hline
	Set A, OBE-sim & 3.64 & 5.1 & 9.5 \\
        Set C, OBE-sim & 3.52 & 5.0 & 9.0 \\
	Set A, FSS-sim & 3.64 & 5.5 & 9.9 \\
   \hline 
  \end{tabular}
\end{table}

\begin{table} \centering \leavevmode 
\caption{The energy change due to quark Pauli 
     effects. The baryon size is set to $b=0.86$ fm. Energy is given in 
     units of MeV.  $\delta E$ denotes the binding energy change 
     $B_{\Lambda}(b=0)-B_{\Lambda}(b=0.86)$ for $_\Lambda^5$He or 
     $B_{\Lambda\Lambda}(b=0)-B_{\Lambda\Lambda}(b=0.86)$ for 
     $_{\Lambda\Lambda}^{\ \ 5}$H. 
   \label{ENEDIFFQP}}
 \begin{tabular}{cccccc}
  \hline 
  &  & 
	$\varepsilon^2$ & $\langle\psi_{\rm F}|H|\psi_{\rm F}\rangle$ 
  & $\varepsilon^2 \langle\psi_{\rm F}|H|\psi_{\rm F}\rangle$ &  $\delta E$ \\ 
  \hline
	\Lower{$_\Lambda^5$He} & set A  & 0.00406 & 513 & 2.1 & 1.34 \\
	& set C & 0.00331 & 969 & 3.2 & 1.37 \\
  \hline
	& set A, OBE-sim & 0.00117 & 827 & 0.97 & 0.5 \\
	$_{\Lambda\Lambda}^{\ \ 5}$H & set A, FSS-sim 
        & 0.00184 & 522 & 0.96 & 0.6 \\
        & set C, OBE-sim & 0.000815 & 1511 & 1.2 & 0.5 \\
  \hline 
 \end{tabular}
\end{table}

\end{document}